\documentclass[aps,pre,reprint,superscriptaddress,amsmath,amssymb]{revtex4-1}
\usepackage{graphicx}% Include figure files
\usepackage{dcolumn}% Align table columns on decimal point
\usepackage{bm}% bold math
\usepackage{hyperref}%
\usepackage{amsmath}
\usepackage{amssymb}
\usepackage{subfigure}
\usepackage{xcolor}

\DeclareMathOperator*{\argmax}{arg\,max}

\begin{document}
\newcommand{\TriDown}{$T^{-}$}
\newcommand{\TriUp}{$T^{+}$}
\newcommand{\AsUp}{$A^{+}$}
\newcommand{\AsDown}{$A^{-}$}

\newcommand{\mat}[1]{{\bf #1}}

%\preprint{}

\title{Network Reliability: The effect of local network structure on diffusive processes}



%\thanks{}


% add \affiliation{Department of Physics, Virginia Tech, Blacksburg, Virginia 24061, USA} for Stephen and Yasamin
\author{Mina Youssef}
\email{myoussef@vbi.vt.edu}
\affiliation{Network Dynamics and Simulation Science Laboratory, Virginia Bioinformatics Institute, Virginia Tech, Blacksburg, Virginia 24061, USA}

\author{Yasamin Khorramzadeh}
\email{yasi@vbi.vt.edu}
\affiliation{Network Dynamics and Simulation Science Laboratory, Virginia Bioinformatics Institute, Virginia Tech, Blacksburg, Virginia 24061, USA}
\affiliation{Department of Physics, Virginia Tech, Blacksburg, Virginia 24061, USA}

\author{Stephen Eubank}
\email{seubank@vbi.vt.edu}
\affiliation{Network Dynamics and Simulation Science Laboratory, Virginia Bioinformatics Institute, Virginia Tech, Blacksburg, Virginia 24061, USA}
\affiliation{Department of Physics, Virginia Tech, Blacksburg, Virginia 24061, USA}
\affiliation{Department of Population Health Sciences, Virginia Tech, Blacksburg, Virginia 24061, USA}

%\date{\today}

\begin{abstract}
%To study the influence of structural properties on the spread of diseases using the network reliability, 
%This paper addresses the influence of structural properties on the spread of diseases using the network reliability polynomial. 

This paper re-introduces the network reliability polynomial -- introduced by Moore and Shannon in 1956 -- for studying the effect of network structure  on the spread of diseases. 
We exhibit a representation of the polynomial that is well-suited for estimation by distributed simulation. 
We describe a collection of graphs derived from Erd\H{o}s-R\'{e}nyi and scale-free-like random graphs in which we have manipulated assortativity-by-degree and the number of triangles. 
We evaluate the network reliability for all these graphs under a reliability rule that is related to the expected size of a connected component.
Through these extensive simulations, we show that for positively or neutrally assortative graphs, swapping edges to increase the number of triangles does not increase the network reliability. 
Also, positively assortative graphs are more reliable than neutral or disassortative graphs with the same number of edges. 
Moreover, we show the combined effect of both assortativity-by-degree and the presence of triangles on the critical point and the size of the smallest subgraph that is reliable. 

\end{abstract}

%\keywords{}

\maketitle

\section{Introduction}

We study the dynamics on a variety of networks for a networked $S-I-R$ model of epidemics, in which each vertex can be in one of the three states $Suscepitble$, $Infectious$, or $Recovered$  \cite{Eubank:04,Eubank:10}. 
%{\color{red} Need a reference to MIDAS}. 
As is well known, this process is equivalent to bond percolation \cite{Grassberger:83}, and thus exhibits a percolation phase transition and associated critical phenomena in an infinite network. 
The mean field dynamics are also well understood: critical phenomena such as scaling exponents depend only on the degree, also known as the coordination number. Corrections to mean field dynamics \cite{Kermack:27, Anderson:92} have been established that take into account variations in degree from one vertex to another \cite{Moreno:02, Youssef:11}. Often, following \cite{Barabasi:99}, the variation is taken to follow a power law distribution. However, the most important variation is not necessarily in degree, but in the number and overlaps of loops of a given length. Both the degree and the distribution of loops are completely determined by the dimension for regular grids, where much of the theory was developed, but not for generic graphs. 
%\cite{loop expansions}
In this paper, we illustrate how to use the concept of network reliability to elucidate how details of network topology influence the spread of epidemics. 
There are many structural aspects of contact networks that interact in complicated ways with each other and with the dynamical properties of disease transmission to create population-level dynamics in infectious disease outbreaks. 
For concreteness, we focus on the effect of degree assortativity and the number of triangles. 
As we show below, the complicated interaction between these structural measures generates a wide range of population-level effects.
% Thus, characterizing contact networks in a way that is directly relevant to epidemiology is important, but difficult. 

We show how to characterize a network by the way its overall attack rate -- the mean cumulative fraction of vertices infected before this transient dynamics reaches a fixed point -- varies with disease transmissibility.
Interventions alter the network structure, changing the overall attack rate. In \cite{Marathe:11}, we found that isolating infected people within a household, i.e.\ limiting their contacts with other household members, can significantly reduce the population-level attack rate for a wide range of transmissibility. 
In this case we can characterize the network after intervention as uniformly more resistant to epidemic outbreak than the original network.  
  
% The impact of network structure on diffusion dynamics can be addressed using 
The overall attack rate is a special case of the \emph{Network Reliability Polynomial}\cite{Moore:56} formalism.
This formalism was introduced to analyze specific networks.
Hence, one of its strengths for characterizing networks is that it makes no assumptions about regularities or symmetries.
We define and provide algorithms for calculating and estimating coefficients of the reliability polynomial, 
% discuss its interpretation in terms of statistical physics, 
provide illustrative examples on several networks, and show how it can be used to understand complicated phenomena.
%We illustrate how to compare two graphs and identify the structural differences between them that are most relevant for a given dynamical process.
% We also indicate how reliability can be used to infer structure, in the sense of network tomography.
The novelty in this work is not the concept of reliability itself -- the {\em IEEE Transactions on Reliability} is now in its 61st year -- nor is it in the statistical physics of reliability.
It is in our suggestions that 
\begin{enumerate}
\item coefficients of the reliability polynomial are the best way to characterize graph structure and
\item network analysis in terms of reliability provides insights into global effects of local structural details that elude other approaches.
\end{enumerate}

Reliability refocuses the question of structural effects from the individual interactions between elements to global dynamical properties, suggesting new methods of analysis.
The coefficients of the reliability polynomial transform all the information in the network adjacency matrix into a form that, by design, 
reflects dynamical phenomena of interest.
Hence it is a structural measure that is immediately connected with dynamics. 
% The reliability coefficients fold together static measures like degree, modularity and measures of centrality, into precisely the combinations that are most relevant to the dynamics. . 
% Moreover, the reliability can be further decomposed into the product of a purely combinatorial factor and a structure-dependent factor.
% Furthermore, network reliability provides information about the diffusion of dynamics as a polynomial $R(x)$. 
Network reliability is amenable to study from many perspectives, and much is known about the general properties of the reliability polynomial \cite{Colbourn}. 

In contrast, the literature about the relation between dynamics and common graph statistics such as assortativity-by-degree and clustering coefficient is confusing and sometimes inconsistent. 
For example, consider what is known about the relationship between the spread of $S-I-R$ epidemics and assortativity-by-degree.
Assortativity can be defined as a correlation coefficient between the degrees of vertices at each end of an edge.
Thus it ranges from highly assortative (+1) through neutrally assortative (near 0) to highly disassortative (-1).
%The spread of $S-I-R$ epidemics in positively and neutrally assortative networks has been addressed in \cite{Moreno:03}. 
The spread of $S-I-R$ epidemics in correlated and uncorrelated networks has been addressed in \cite{Moreno:03,Nold:80,Newman:02,Kiss:08,Agostino:12,Badham:10}. 
%{\color{red} also cite the other references in this paragraph here, e.g. Nold, Newman, Kiss, Agostino, Badham ?}
Given a social network, Nold in \cite{Nold:80} grouped individuals based on their number of contacts. Thus, high epidemic prevalence appears in groups with highest number of contacts.  
In contrast to Nold, Moreno and Pacheco\cite{Moreno:03} reported that positively assortative networks have fewer large outbreaks than neutrally assortative networks. 
Moreover, for finite size networks, the epidemic threshold for positively assortative networks is larger than that for neutrally assortative networks, indicating more robustness against the spread of epidemics. 
Consistent with Newman\cite{Newman:02}, epidemics persist in positively assortative networks longer than in neutrally assortative networks when the initial infected vertex is the one with the largest node degree. 
Kiss and Kao\cite{Kiss:08} showed that epidemics spread faster in positively assortative networks than in disassortative (negatively assortative) networks. 
However, this result disagrees with D'Agostino et al.\cite{Agostino:12}, in which it is shown that disassortative networks have a shorter longest time to peak epidemic prevalence than assortative networks. 
The longest timescale is the inverse of the algebraic connectivity representing the slowest mode of diffusion in the network \cite{Almendral:07}.
Disassortative networks have larger algebraic connectivity than assortative networks.
 In other words, disassortative networks have shorter longest timescale to the epidemic peak than assortative networks. Thus, epidemics spread faster in disassortative networks than in assortative networks. 
 % Consequently, interventions are more effective in assortative networks than in disassortative networks. 
 Finally, the combined impact of both assortativity-by-degree and clustering coefficient on the spread of epidemics is studied in Badham and Stocker\cite{Badham:10}. Through extensive simulations on a limited set of networks, the authors found that both the total epidemic size and the average secondary infection size are smaller for highly clustered and/or highly positively assortative networks. 
 However, for smaller values of these properties, the epidemic final size is inconsistent with the increase of either the assortativity value or the clustering coefficient.

% derive two equivalent representations of reliability, each appropriate for different kinds of problems, and 
%The outline of this paper is as follows: We re-introduce network reliability and its relationship to statistical physics. Then we 
The outline of this paper is as follows: First, we re-introduce network reliability in terms of reliability rules and reliability polynomials. Then we 
discuss the estimation of reliability coefficients. We describe an {\em in silico} laboratory of networks with a range of carefully controlled topological properties. We  characterize these networks' reliability in terms of critical points and other features, elucidating the relationship between network reliability and common graph statistics as a function of network size.  Finally, we indicate some intriguing open research problems.  

\section{Network reliability}

See Colbourn \cite{Colbourn} for a comprehensive introduction to notions of reliability. Consider a graph $G(V,E)$ with $V$ vertices and $E$ weighted edges.
The edges may be directed or undirected, and there may be multiple edges between two vertices.
Let the set ${\cal S}$ be the set of all subgraphs of $G$ generated by including each edge $(i,j)$ independently with probability $x_{i,j}$.
There are $2^E$ elements of this set.

Now consider a binary function $r:{\cal S}\rightarrow \{0,1\}$, the {\em reliability rule}.
If $r(s) = 1$, we say that subgraph $s$ is {\em accepted} or {\em reliable}.
We define the {\em reliability} $R(G, r, \{x\})$ of a base graph $G$ with respect to the reliability rule $r$ for edge weights $\{x\}$ as the probability that a randomly chosen subgraph $s$ is reliable.
In other words, a network is reliable to the extent that it remains functional under random removal of edges.
Formally:
\begin{equation}
R(G, r, x) \equiv \sum_{s \in {\cal S}} r(s) p_x(s).
\label{eq:defn}
\end{equation}
We will explicitly include the dependence on the graph $G$ and the rule $r$ in notation such as $R(G,r,x)$  only when we wish to distinguish the reliability of two different graphs or two different rules.

\subsection{Reliability rules}

There are many useful reliability rules, for example:
\begin{enumerate}
\item {\em two terminal:} a subgraph is accepted if it contains at least one directed path from a distinguished vertex $S$ (the {\em source}) to another distinguished vertex $T$ (the {\em terminus});
\item {\em at-least-n-terminal:} a subgraph is accepted if it contains at least one connected component of size $n$ or greater;
\item {\em all-terminal:} a subgraph is accepted if it is connected and contains every vertex of the base graph;
% \item {\em percolating:} a subgraph is accepted if it contains a path from a source to at least one of the vertices at maximum distance from that source, i.e.\ the ``other side'' of the network;
% \item {\em motif:} a subgraph is accepted if it contains at least one set of edges arranged as in a particular motif, e.g.\ triangle, star.
\item {\em attack rate (AR)-$\alpha$:} a subgraph is accepted if the mean component size across all {\em vertices} is greater than or equal to $\alpha V$. Note that this is different from the mean component size taken across all {\em components}. In fact, it is the sum taken across all {\em components} of the {\em squared} component size divided by $V$.
\end{enumerate}
For graphs with directed edges, the notion of ``connected'' can be generalized appropriately.
We primarily use the AR-$\alpha$ rule in this paper because of its epidemiological relevance. 
As its name suggests, it gives the probability that the cumulative fraction of vertices infected (sometimes called the ``wet set'' in a percolation setting) exceeds $\alpha$, averaged over all possible initial conditions in which a single vertex is infected.
This rule, along with many other commonly used rules, has the useful property of {\em coherence}, i.e.\ adding an edge to a reliable subgraph does not make it unreliable.

%It is important to note that the arguments below do not depend on the form of the reliability rule. 
%For example, a reliability rule requiring that a connected component be {\em exactly} size $n$ would not be coherent, but one that substitutes size $\ge n$ would.
%In principle, it is possible to add constraints on subgraph statistics to the reliability rule, 
%e.g.\ that a subgraph's clustering coefficient must lie in a range of values.
%However, the resulting rule is no longer necessarily coherent rendering it difficult to analyze in this framework.
%We discuss in Section~\ref{sec:Bayesian} an alternative way to obtain the same information.

\subsection{Reliability polynomials}

The reliability defined in Equation~\ref{eq:defn} depends on the probability of obtaining each particular subgraph when edges are selected {\em independently} at random.
It is this independence of selecting edges that makes reliability such a powerful tool.
For instance, the probability of selecting any particular subgraph is simply the product of the probability of selecting each of its edges and not selecting each edge that doesn't appear.
As we show below, when the edges are homogeneously weighted with, say, uniform probability of selection $x$, this reduces to a  homogeneous polynomial in $x$ and $(1-x)$ of degree $E$.
The case of a few different weights can either be treated by considering a multivariate polynomial in the weights, or by adding multiple edges between vertices;
for many different weights, this becomes intractable.
We restrict ourselves to the homogeneously weighted case in this paper.

To rewrite Equation~\ref{eq:defn} in polynomial form, first partition the set of subgraphs ${\cal S}$ into subsets ${\cal S}_k$ in which each subgraph has exactly $k \le E$ edges.
Each subgraph with $k$ edges appears with probability $p=x^k(1-x)^{E-k}$.
If ${\cal R}$ denotes the set of all reliable subgraphs, then
$R_k \equiv |{\cal R} \cap {\cal S}_k|$ is the number of subgraphs with exactly $k$ edges that are accepted by rule $r$.
Then the total contribution of subgraphs in ${\cal S}_k$ to $R(x)$ is $R_k x^k(1-x)^{E-k}$.
Summing these contributions over all $k$ gives the {\em reliability polynomial} (for rule $r$ and graph $G$):
\begin{equation}
R(x) = \sum_{k=0}^E R_k x^k(1-x)^{E-k}.
\label{eq:xpoly}
\end{equation}

\begin{figure}
\includegraphics[width=9cm]{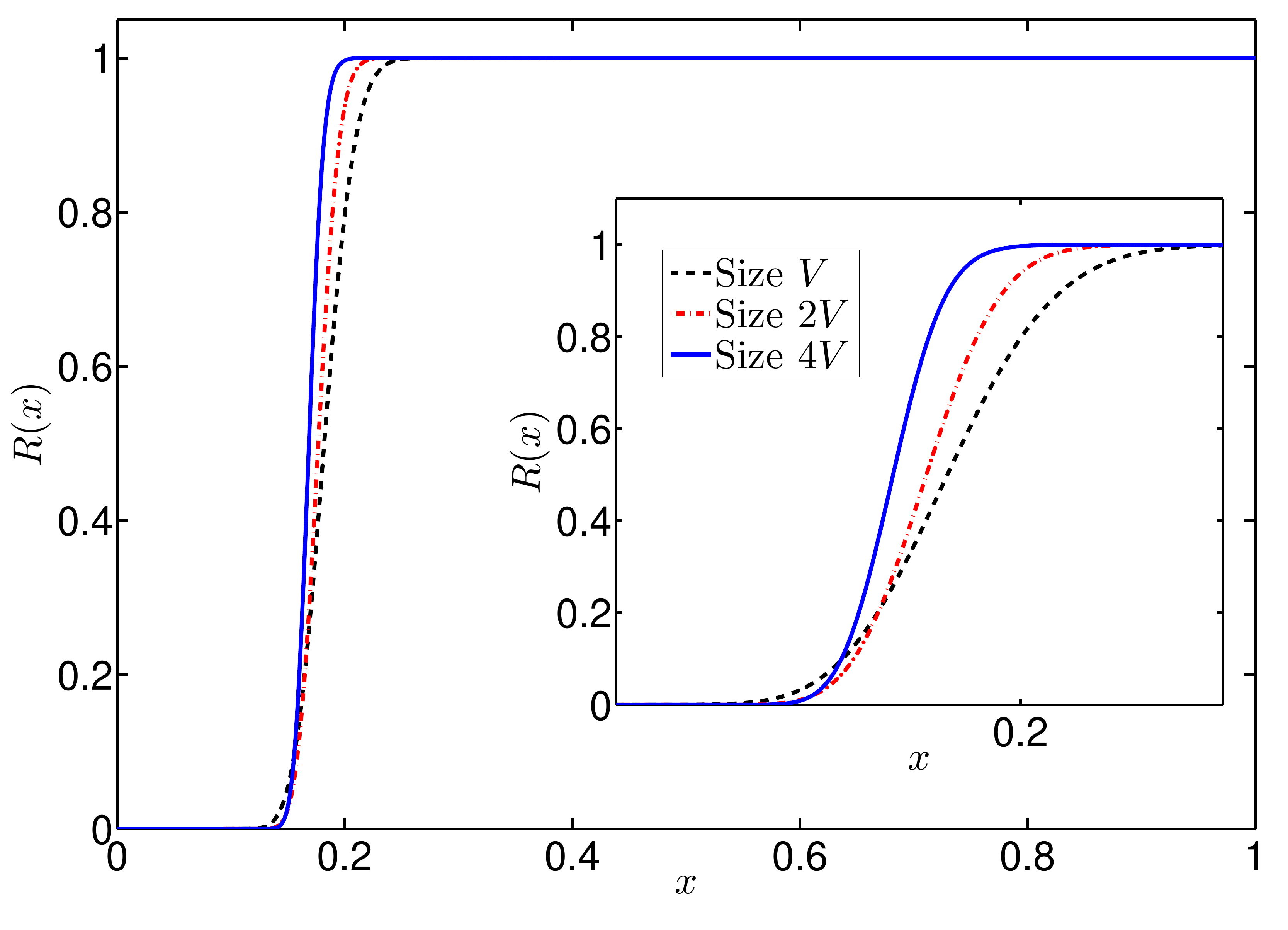}
\caption{Network reliability $R(x)$ for assortative Erd\H{o}s-R\'{e}nyi GNM graphs with sizes $V$, $2V$ and $4V$. The inset shows the transition from $R(x)=0$ to $R(x)=1$.}
\label{fig:GNMrx}
\end{figure}

Figure \ref{fig:GNMrx} shows the network reliability $R(x)$ for several Erd\H{o}s-R\'{e}nyi GNM graphs that have been rewired to have positive assortativity-by-degree. 
General properties to note are: $R_k$ is a non-negative integer in the range $[0, {E \choose k}]$; $R(0) = 0$, $R(1) = 1$ for non-trivial networks; and for a coherent rule, $R$ is monotonic non-decreasing.

We can rewrite $R_{k}$ as a product of two factors, taking
\begin{equation}
R_k = P_k {E \choose k}
\label{eq:Pdefn}
\end{equation}
as a definition of $P_k$. This decomposition splits $R_k$ into what we might call an {\em entropic} or {\em combinatorial} factor ${E \choose k}$ and a {\em structural} factor $P_k$.
The entropic factor simply makes explicit the sharp peak in the number of possible subgraphs with $k$ edges around a small region centered at $k = Ex$, i.e.\ the size of the space from which equi-probable system configurations can be drawn.
%It creates an envelope that windows the effects of $y^k$ to a small region centered at $k = Ex$.
The factor $P_k$ is structural in the sense that it encodes all the information about the specific graph $G$ that is needed to determine its reliability. 
%This is obvious because the probability $y^k$ and the entropic factor $E \choose k$ are the same for any graph with $E$ edges.
% In Appendix A, we present constraints on $R_k$ and $P_{k}$.

The interpretation of $P_k$ is clear -- it is the fraction of possible subgraphs with $k$ edges that are accepted by the reliability criterion. 
This interpretation suggests a simple estimation procedure for
$P_k$: select a sample of subgraphs with $k$ edges, evaluate the reliability criterion for each, and let the estimated $P_k$ be the fraction of the sampled subgraphs that are reliable. 
Given a graph in memory, the computational complexity of selecting a subgraph is proportional to $k$ (not $E$), the number of samples selected, and the complexity of evaluating the criterion.
(The complexity of the criterion itself should not be overlooked. For most reliability rules discussed here, it can be evaluated by partitioning the selected subgraph into connected components.)
Moreover, since each subgraph can be chosen and its reliability evaluated independently, the algorithm can be distributed easily onto massively parallel distributed machines.

\subsection{Alternative expressions for $R(x)$}

There are many possible complete sets of basis functions for polynomials on the unit interval in general, and hence for the reliability polynomial in particular.
We find two to be particularly useful, even though they are not orthogonal bases: 
\begin{enumerate}
% \item a function of the continuous variable $x$, i.e.\ the value of the polynomial, $R(x)$;
\item the set of $E$ functions ${E \choose k} x^k(1-x)^{E-k}$.
% where $k$ is the number of edges in a subgraph.
The coefficients in this basis are the $P_k$ introduced above.
Although, as discussed in Colbourn, evaluating these coefficients exactly is computationally hard for many graphs and many reliability rules, the
$P_k$ can be estimated to arbitrary precision by a simple, scalable algorithm for any graph.
This basis is thus well-suited for computational analysis of particular graphs.
\item the set of $E$ functions $x^k$. 
The coefficients in this basis, which we denote by $N_k$, can obviously be drives from the $P_k$ by expanding the binomial $(1-x)^{E-k}$, but as we show in a companion manuscript, they also have an important physical interpretation in terms of the number and overlaps of what we call {\em structural motifs}.
This basis is well-suited for reasoning about graph structure in general.
\end{enumerate}

\section{Important features of network reliability}

\subsection{Minimum and maximum number of edges of reliable subgraphs}

In Figure~\ref{fig:GNMrx}, note that $R(x)$ is negligible for $x <0.1$ and is near unity for $x > 0.25$, i.e.\ for subgraphs with fewer than $k=0.1E$ edges or more than $0.25E$ edges, respectively.
This is a common feature of network reliability for many different rules, related to max flow / min cut theorems.
Let $k_{min}+1$ represent the minimum number of edges for any subgraph to be reliable, so that $P_{k_{min}}=0$ and $P_{k_{min}+1}>0$. 
Similarly, let $k_{max}$ represent the minimum number of edges that are necessary for every subgraph to be reliable, so that $P_{k_{max}-1} < 1$ and  $P_{k_{max}}=1$. It is always true that $k_{min} < k_{max}$ since the reliability rule is coherent, i.e.\ adding an edge to a reliable subgraph does not make it unreliable. 
Thus we can write the probability $P_{k}$ as follows:

\begin{equation}
P_k = \left\{
\begin{array}{lr}
0, & k \leq k_{min} \\
0 < P_{k} < 1, \qquad & k_{min} < k < k_{max}\\
1, & k_{max} \leq k \leq E.
\end{array}
\right.
\end{equation}

\subsection{Average reliability}

The average reliability $\langle R(x) \rangle$ 
%vrepresents the projection of the function $R(x)$ onto its scalar mean value. It 
gives the expected outcome for a disease with unknown transmissibility \cite{Youssef:11b,Mieghem:12b}. 
The transformation between reliability $R(x)$, viewed as a function of $x$, and its coefficients $P_k$, viewed as a function of $k$, has the following nice property:
 the average value of $R(x)$ is equal to the average value of $P_k$.
To demonstrate this, first note that the recursion formula:
\begin{eqnarray}
h(a,b) &\equiv& \int_0^1 x^a(1-x)^b dx  \nonumber \\
 &=& (a+1)^{-1} 
\left\{
\begin{array}{cc}
1  & b=0   \\
b h(a+1, b-1) & b>0  
\end{array}
\right.
\end{eqnarray}
has the solution
\begin{equation}
h(a,b) = \frac{a! b!}{(1+a+b)!}.
\end{equation}

Then interchanging the sum and the integral and integrating by parts yields
\begin{eqnarray}
\label{eq:averxpk}
\langle R(x) \rangle &\equiv& \int_{x=0}^{x=1}\sum_{k=0}^{E} R_{k} x^{k} (1-x)^{E-k} dx \nonumber \\
&=& \sum_{k=0}^{E} P_{k} {E \choose k} \frac{k! (E-k)!}{(1+E)!} \nonumber \\
&=& (1+E)^{-1} \sum_{k=0}^{E} P_{k} \nonumber \\
 &\equiv& \langle P_{k} \rangle
\end{eqnarray}

Thus, the average reliability represents the average probability of selecting a reliable subgraph.

\subsection{Critical point}

Equations~\ref{eq:defn} or \ref{eq:xpoly} define a partition function for the system: the weighted sum over all configurations (subgraphs) of the reliability of the configuration, weighted by its probability.
Thus the reliability can be viewed as an order parameter for the system. 
In the thermodynamic limit, i.e.\ for an infinite graph, we expect that the derivative of the reliability with respect to $x$ will diverge at a {\em critical point $x_{c}$} of a phase transition, for example, the percolation phase transition for the all-terminal reliability rule.
For finite graphs, we take the value of $x$ for which the derivative of the reliability attains its maximum as defining the critical point $x_{c}$.
The first derivative of reliability is the probability that the reliable subgraph starts to percolate if the probability of choosing an edge increases from $x$ to $x+dx$~\cite{Stauffer:91}. 
From Equations \ref{eq:xpoly} and \ref{eq:Pdefn}, we find that the first derivative of the reliability can also be written as a homogenous polynomial in $x$ and $(1-x)$, where the coefficients are finite differences of the $P_k$:

\begin{eqnarray}
\frac{d R(x)}{dx} = \qquad \qquad \qquad & \nonumber \\
  \sum_{k=0}^{E-1} \left[ kP_{k+1} -(k+1)P_k \right] &{E \choose k+1} x^k(1-x)^{E-1-k} 
\end{eqnarray}
%Explicit form of derivative, interpretation in Ising terms

\section{A laboratory for studying graphs}

We have constructed a set of graphs of three different sizes with several different degree distributions but similar mean degree and carefully controlled ranges of assortativity-by-degree
 and number of triangles.
 (For convenience below, we use the general term ``assortativity'' to mean specifically assortativity-by-degree.)
These graphs form an {\em in silico} laboratory for studying structural effects in graphs.
This laboratory, along with software for evaluating network reliability, will be made accessible to the public via the Cyber-Infrastructure for Network Science (CINET) web site~\url{http://ndssl.vbi.vt.edu/cinet}.

\begin{figure*}%[htbp]
\begin{center}
%\vspace{.2in}
%\centerline {
\includegraphics[width=7cm]{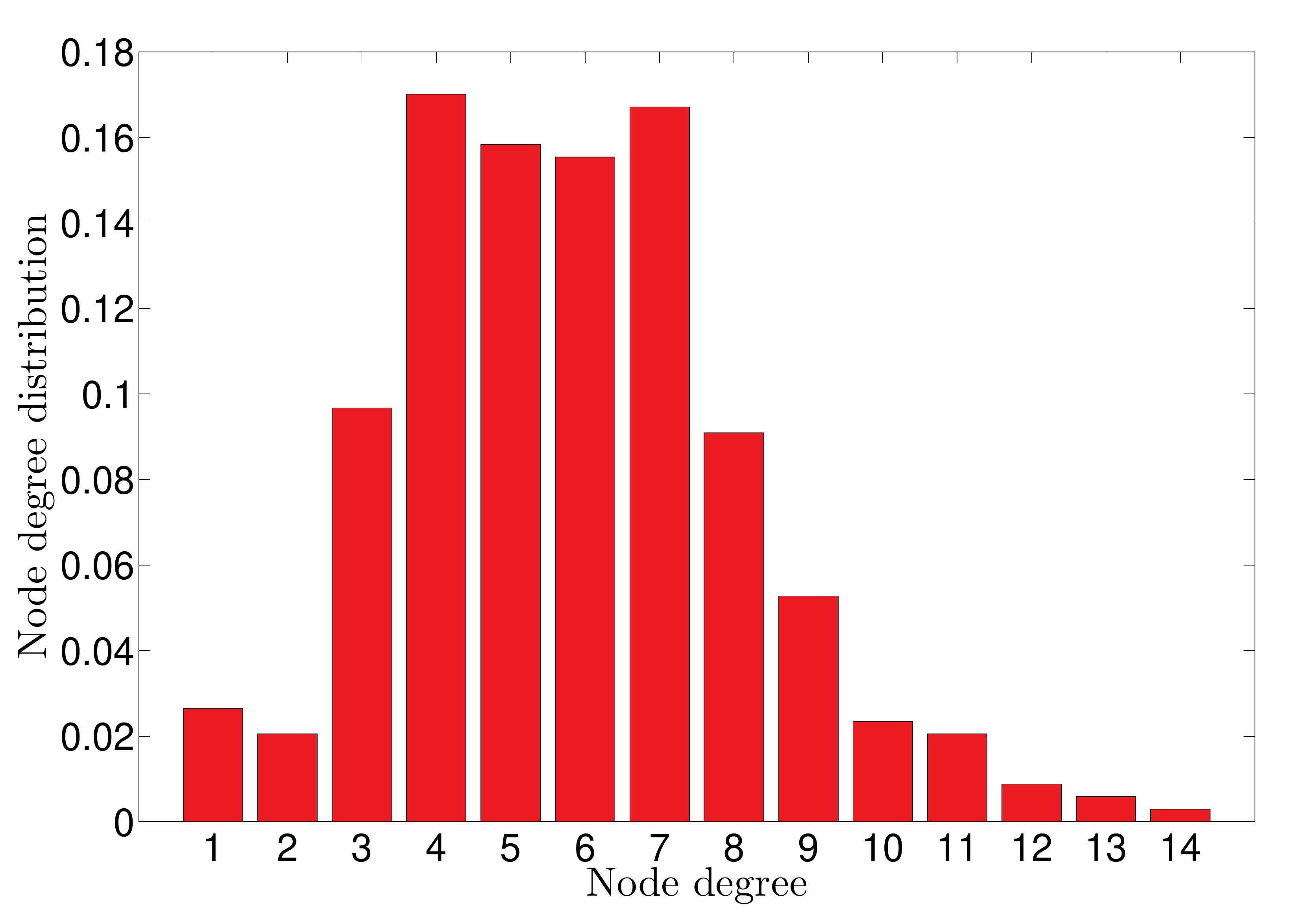}
\includegraphics[width=7cm]{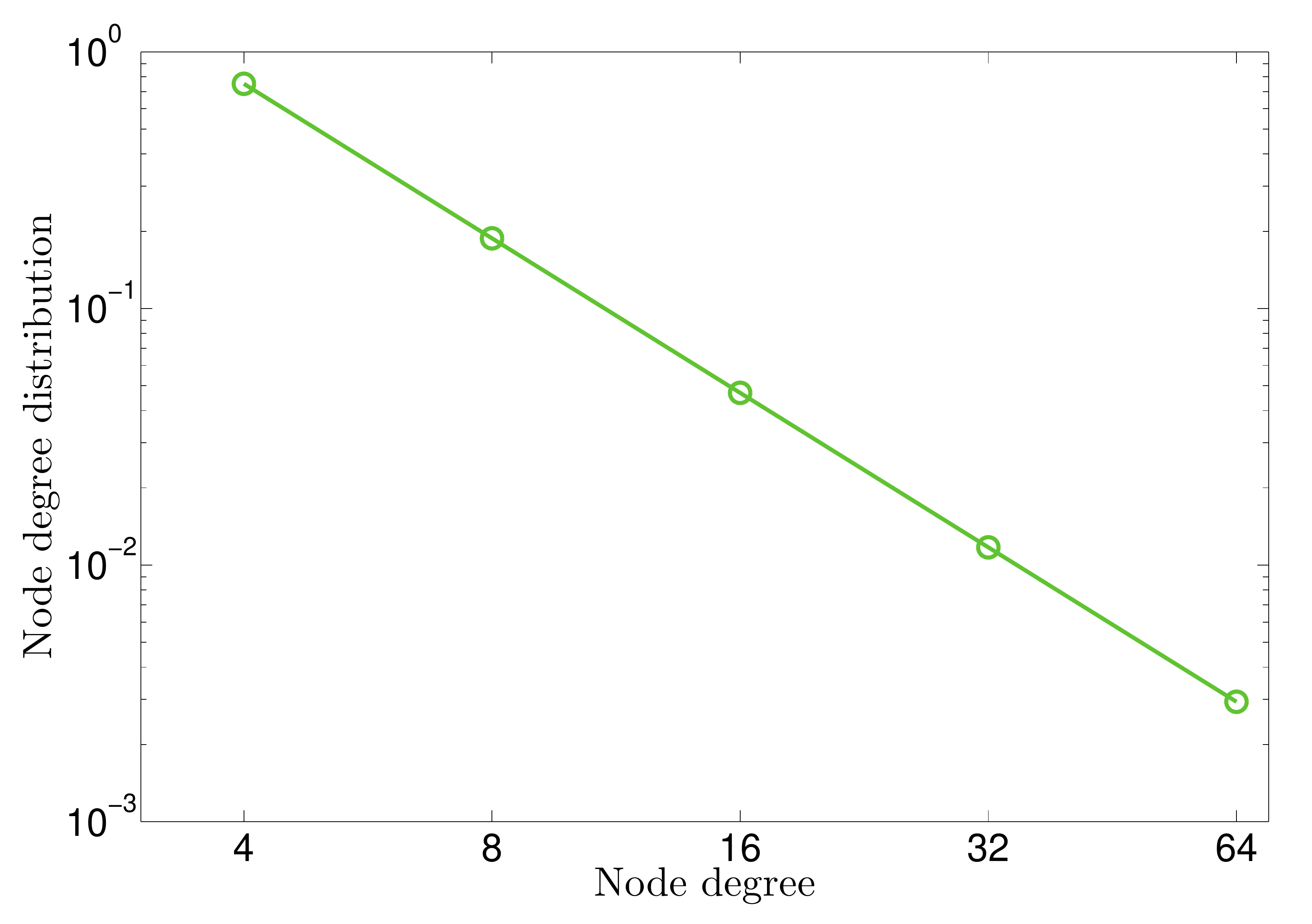}
%}
%\vspace{.2in}
\caption{Degree distributions for the Erd\H{o}s-R\'{e}nyi graphs, GNM (left panel); and the scale-free like graphs, SFL (right panel). Note the logarithimic axes for the SFL graphs.}
\label{fig:DD}
\end{center}
\end{figure*}

Beginning with a single randomly generated graph instance for each of two degree distributions, we apply assortativity and triangle ``raising and lowering'' operators $A^\pm$ and $T^\pm$ defined as follows:
\begin{itemize}
\item The $A^{+}$ and $A^{-}$ operators increase or decrease, respectively, a graph's assortativity-by-degree.
\item The $T^{+}$ and $T^{-}$ operators increase or decrease, respectively, the number of triangles in a graph while leaving its assortativity-by-degree invariant.
\end{itemize}
%Since the operators we use to manipulate assortativity and clustering maintain the degree distribution invariant, the real difference among the manipulated graphs is in their degree distribution. 
The first graph is an Erd\H{o}s-R\'{e}nyi random graph (or $GNM$ for short) -- i.e.\ one generated by choosing $E$ edges uniformly at random from among $V$ vertices, with $V=341$ and $E=992$ in this case.
The reason for choosing these values of $V$ and $E$ will become clear below.
The degree distribution of this graph is as follows, illustrated in Figure~\ref{fig:DD}a:
$(1, 9), (2, 7), (3, 33), (4, 58), (5, 54), (6, 53), (7, 57), (8, 31),$ $(9 18), (10, 8), (11, 7), (12, 3), (13, 2), (14, 1))$.

We accepted the first generated $GNM$ that was also connected.
We claim that this bias toward connectivity has not produced an atypical degree distribution.
In the limit as $E \to \infty$ with fixed $E/V$, the expected degree distribution becomes Poisson, as is well known.
Note, however, that the expected degree distribution of {\em connected} $G(V,E)$ is slightly different from that of {\em all} $G(V,E)$, since it is less likely that a graph with many vertices of low degree is connected.
Consider, for example, that a connected graph cannot have any vertices with degree 0.
Selecting any vertex as part of an edge is a Bernoulli process with probability ${2 \over V}$.
Hence across all $G(V,E)$, the probability of observing a vertex with degree $d$ is $p(d) = {E \choose d} \left({2 \over V}\right)^d \left(1 - {2 \over V}\right)^{E-d}$.
%For $N=341$ and $m=992$, the expected degree distribution and the degree distribution of our connected $G(N,m)$ are shown in Figure~\ref{fig:DD}a.
Thus, roughly 37\% of all $G(V=341,E=992)$ will have no vertices with degree 0.
While this condition alone is not a guarantee of connectedness, it indicates that the degree distribution for our sample graph is not atypical.
%The probability of observing any particular degree distribution is $\prod_{d=0}^{N-1} p(d)$.
%Thus the probability of observing the expected degree distribution is ...;
%the probability of observing the degree distribution of our connected $G(N,m)$, on the other hand, is ...

Because of the recent interest in scale free graphs, we also considered a ``scale-free-like'' ($SFL$) graph.
The degree distribution of this graph is as follows, illustrated in Figure~\ref{fig:DD}b:
$(4, 256), (8, 64), (16, 16), (32, 4), (64, 1)$, with, therefore $V=341$ vertices and $E=992$ edges.
We consider it scale-free-like because the frequency of finding a vertex with degree $d$, for those degrees that are present, scales as $d^{-2}$.
% Of course, it is a finite graph and many degrees do not appear in the degree sequence, so it could be argued that it is not truly scale-free.
% Nonetheless, it is a non-trivial graph, with V=341 vertices and E=992 edges.
We have not included vertices of degree 1 or 2 in this graph because they are less interesting dynamically than those of higher degree.

Obviously, the mean degree for the two graphs is the same.
This portion of the CINET graph library includes graphs with several other topologies and degree distributions, e.g.\ regular grids, of nearly the same size and mean degree.

\subsection{Choosing a range of assortativities}
We use the definition of assortativity presented in Newman~\cite{Newman:03}.
We repeatedly apply \AsUp and \AsDown to the instances of $GNM$ and $SFL$ graphs. The operators \AsUp and \AsDown are described in the Appendix.
Applied to GNM, this creates graphs with assortativities in the range $[-0.950, 0.979]$, nearly the full possible range;
for SFL, in the range $[-0.268, 0.248]$, only about one quarter of the possible range and apparently in agreement with an estimate by Newman.

\begin{figure}
\includegraphics[width=9cm]{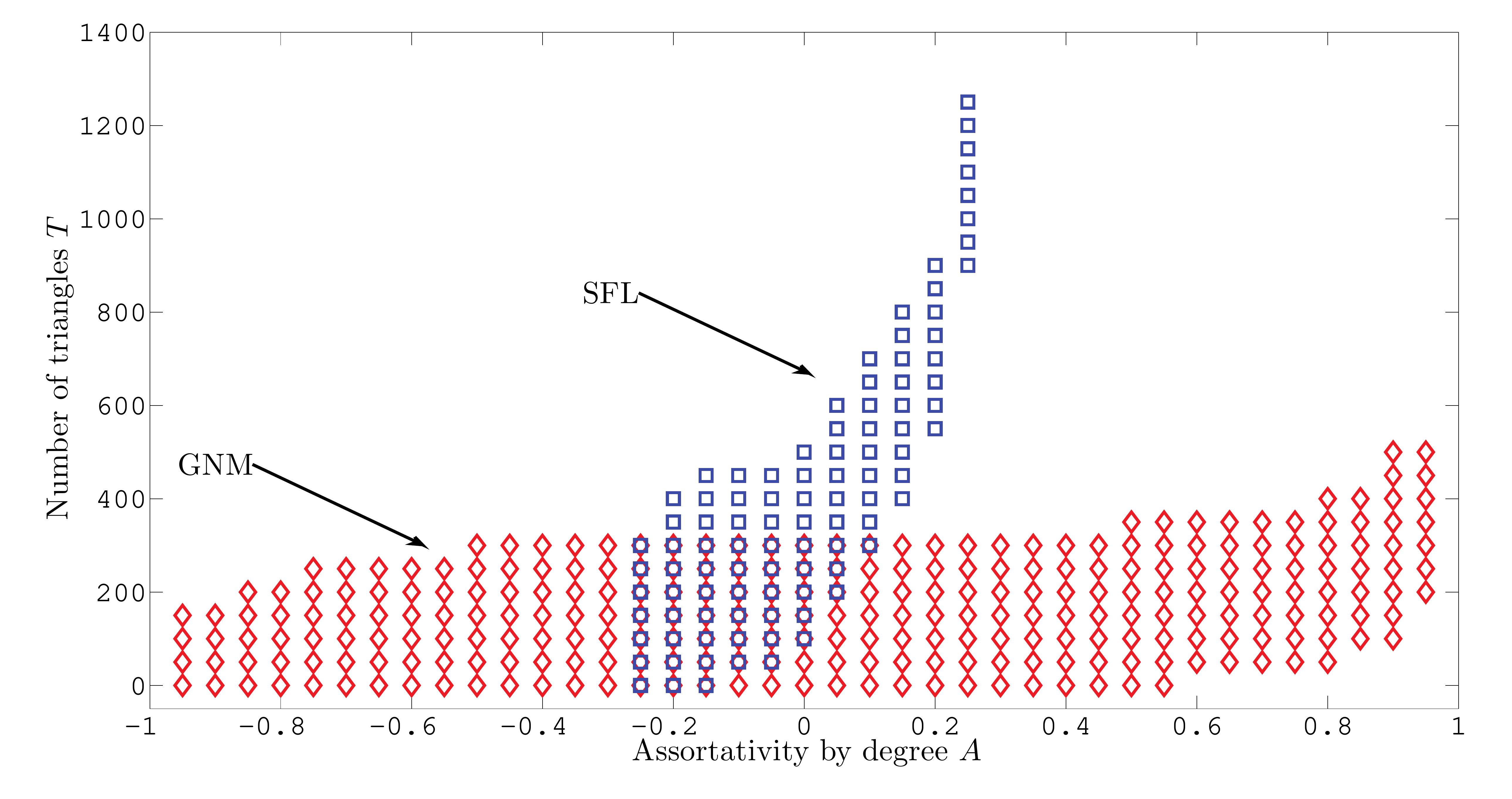}
\caption{Accessible ranges of assortativity and number of triangles for GNM and SFL graphs.}
\label{fig:ExptGrid}
\end{figure}

We  selected for further study only those graphs with assortativities spaced at intervals of approximately 0.05: 41 GNM graphs and 11 SFL graphs.

\subsection{Choosing the number of triangles}
For each value of assortativity, for each degree distribution, we repeatedly apply \TriUp and \TriDown as shown in the Appendix.
The possible range of the number of triangles varies significantly across assortativities and across degree distributions, and is illustrated in Figure~\ref{fig:ExptGrid}.
We chose to study graphs containing approximately multiples of 50 triangles.
%(Most graphs are within 1 triangle of a multiple of 50, a few are 2 away, and the minimum and maximum number of triangles for each assortativity may be off by up to 10.)
%For each of the two initial graphs, Figure~\ref{fig:graphs} displays the initial graph, the version with the highest assortativity and number of triangles, and the version with the lowest assortativity and number of triangles.
Figure~\ref{fig:ExptGrid} shows the locations of all 300+ graphs included in this study in the assortativity - triangles plane.
Clearly, the total number of edges, the degree distribution, and the assortativity place complicated constraints on the total number of triangles in the graph.

Assortativity is defined as a Pearson correlation coefficient, and is thus normalized to lie in the interval $[-1,1]$.
The clustering coefficient for a given vertex $i$ -- and its mean value across all vertices -- can similarly be normalized to lie in the interval $[0,1]$ by dividing the number of triangles including $i$ by the maximum possible number of triangles that could include it, ${d_i \choose 2}$.
However, the value of the clustering coefficient for a graph with a given number of triangles depends on how those triangles are distributed across vertices of different degrees.
Since we explicitly manipulate the assortativity, this distribution changes dramatically.
For example, all else being equal, it is more likely to find a triangle including two vertices of high degree, given that the two are both neighbors of a third.
But all else is not equal -- if the graph is assortative, it is even more likely than if it is disassortative.
For these reasons, in this paper we restrict ourselves to studying the number of triangles directly rather than any normalized version such as the clustering coefficient.

%Thus we do not normalize the number of triangles.
%, defined as the sum over all vertices of the number of triangles including that vertex normalized by ${d^v \choose 2}$, the maximum possible number of triangles that could include that vertex.

\subsection{Choosing the number of vertices}

To study finite size scaling, we constructed graphs with $2V$ and $4V$ vertices.
Since the model used to create the original graphs is specific to the number of vertices, there is some latitude in specifying what it means to scale the number of vertices while maintaining the ``same'' structure.
Specifically, we maintained the edge density (the ratio between number of edges and number of vertices) and the node degree distribution.
%We chose ...

\section{Numerical evaluation of reliability}

% We generate networks with different numbers of triangles and degree assortativity values to 
We evaluated the network reliability for the AR-$\alpha$ reliability rule on all the graphs described in the previous section. 
% All the graphs used in the simulation are connected. 
Recall that AR-$\alpha$ gives the probability that the cumulative fraction of vertices infected exceeds $\alpha$, averaged over all possible initial conditions in which a single vertex is infected. For relevance to the spread of epidemics, we chose $\alpha$ equal 0.2. 

\subsection{Erd\H{o}s-R\'{e}nyi graphs}

\subsubsection{Evaluation of $k_{min}$ and $k_{max}$}

\begin{figure*}
%\begin{center}
\subfigure[GNM graphs]{\label{GNMkminkmax}\includegraphics[height=5.5cm,width=8.0cm]{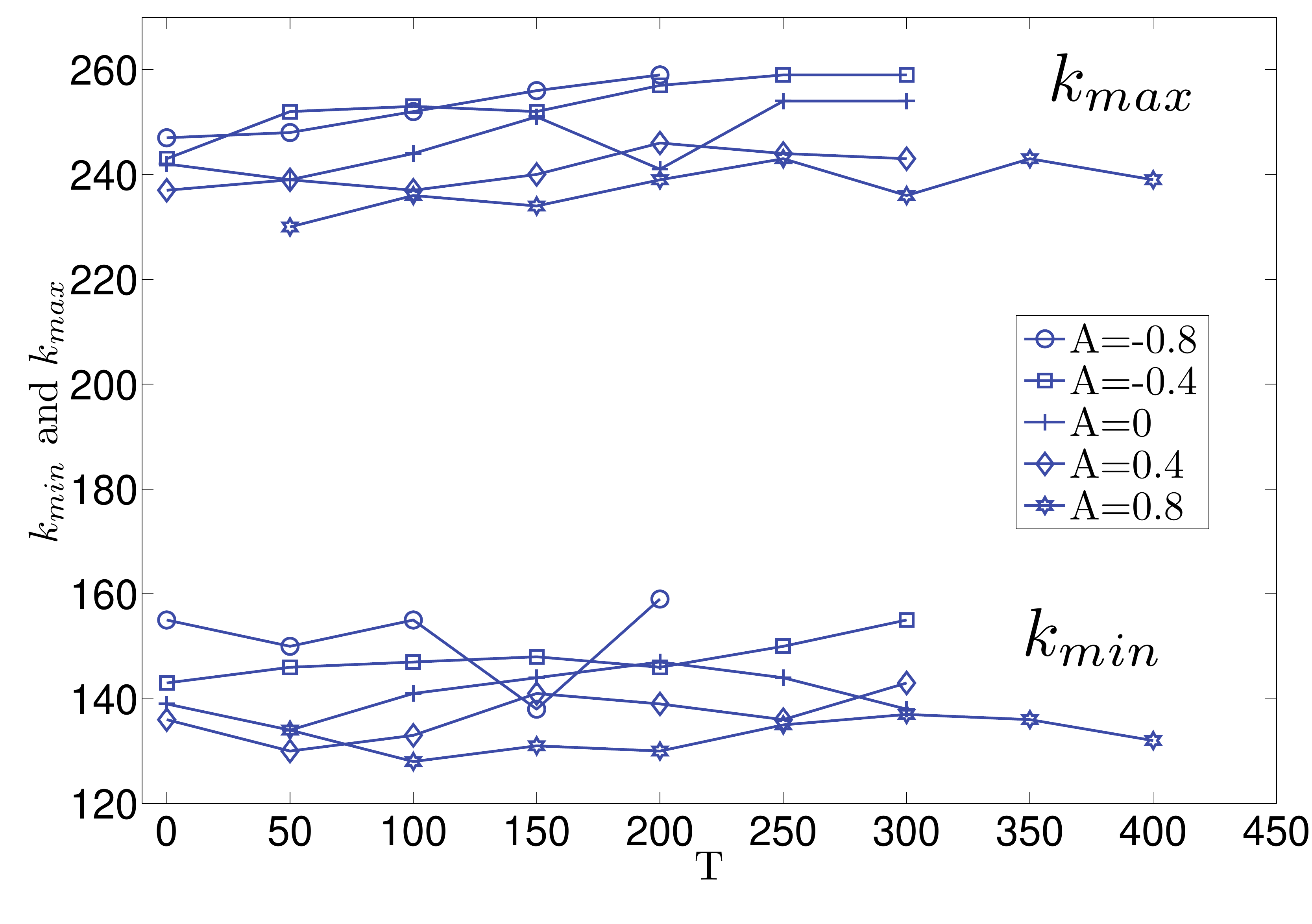}}
\subfigure[SFL graphs]{\label{SFLkminkmax}\includegraphics[height=5.5cm,width=8.0cm]{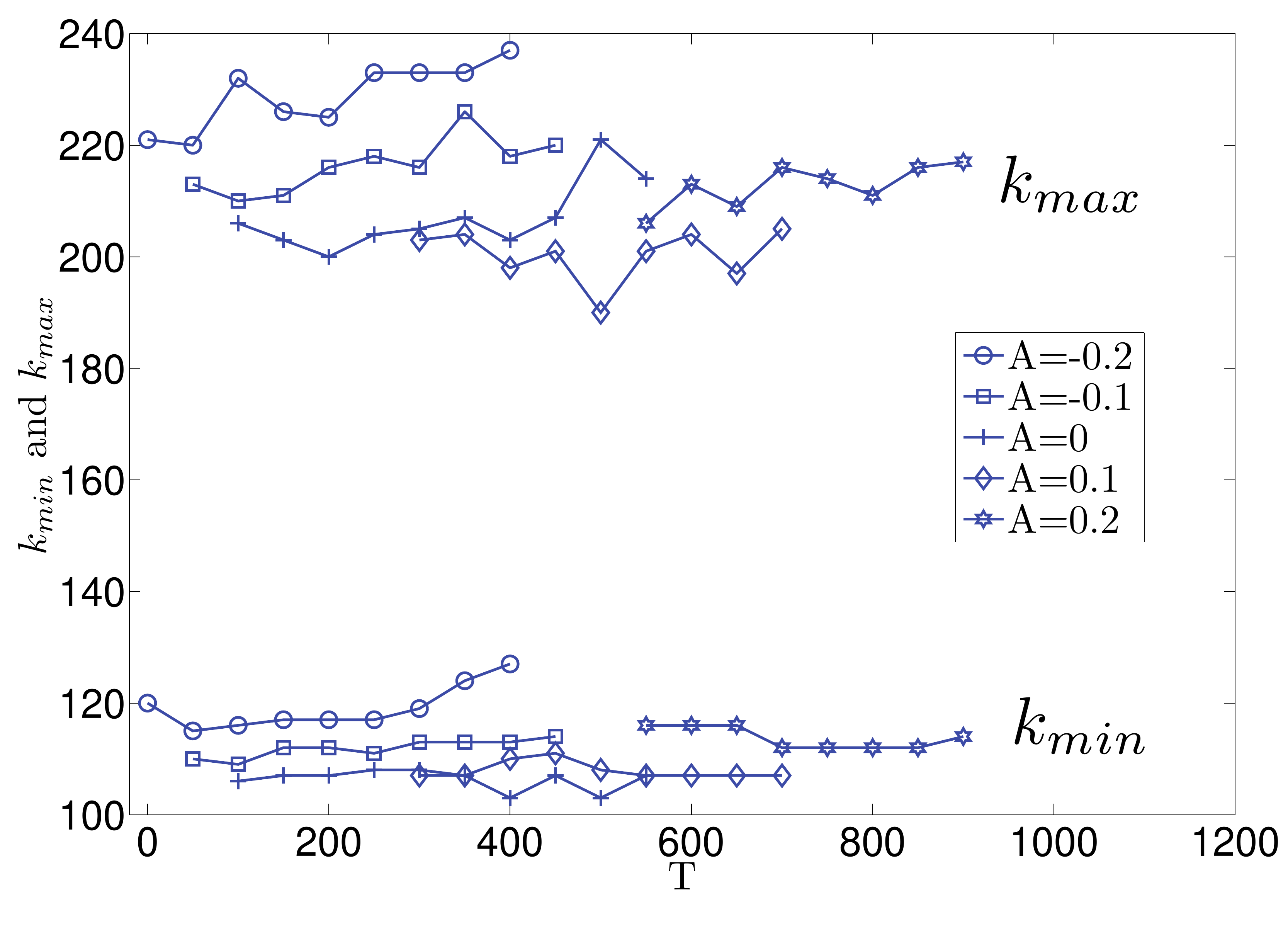}}
%\subfigure[$k_{max}$ for disassortative GNM graphs $A<0$]{\label{}\includegraphics[height=5.5cm,width=8.0cm]{.pdf}}
%\subfigure[$k_{max}$ for neutral and assortative GNM graphs $A\geq 0$]{\label{}\includegraphics[height=5.5cm,width=8.0cm]{.pdf}}
%\end{center}
\caption{$k_{min}$ (bottom) and $k_{max}$ (top) for disassortative, neutral and assortative GNM and SFL graphs under an AR-$\alpha$ reliability rule with $\alpha$=0.2.}
\label{fig:GNMSFLkminkmax}
\end{figure*}

Figure \ref{GNMkminkmax} shows the minimum and maximum number of edges ($k_{min}$ and $k_{max}$) needed to obtain reliable subgraphs for GNM graphs. We observe that, in general, both $k_{min}$ and $k_{max}$ decrease as the assortativity increases. Because $k_{min}+1$ is the minimum number of edges needed to obtain a connected component containing 20\% of the vertices, it represents the edge density of reliable subgraphs. 
Consequently, the edge density of reliable subgraphs is lower for assortative graphs than for neutral and disassortative graphs. 
As mentioned in \cite{Newman:03}, high degree vertices in assortative networks tend to form cliques, which are also called \emph{core groups} in the epidemiological literature. The edge density within the clique is higher than that of the network as a whole. Therefore, a reliable subgraph will first appear with fewer edges within the clique. In disassortative networks, edges tend to connect vertices with dissimilar node degrees. Thus, a reliable subgraph from a disassortative network will first appear with more edges. In other words, reliable subgraphs in assortative networks have lower edge density than reliable subgraphs in disassortative networks as shown in Figure \ref{GNMkminkmax}.
We also observe that the number of triangles has more effect on $k_{max}$ than on $k_{min}$. The number of edges $k_{max}$ increases slightly as the number of triangles increases. 
%Results obtained from SFL graphs are in agreement with results from GNM graphs except when assortativity $A>0.1$, where both $k_{min}$ and $k_{max}$ increase as the assortativity increases, as shown in Figure \ref{SFLkminkmax}. The edge density of reliable subgraphs is larger for SFL graphs with assortativity $A>0.1$ than for neutral SFL graphs. In the next subsection, we explain the effect of high assortativity on the reliability of SFL graphs. 

\subsubsection{Evaluation of critical point and the maximum derivative of reliability}

Figures~\ref{GNMcpA} and \ref{GNMcpB} show the critical point $x_{c}$ for disassortative, neutral and assortative GNM networks. The critical point decreases as assortativity increases; however, the critical point increases as the number of triangles increases. More edges are required to obtain reliable subgraphs from highly clustered graphs. In addition, reliable subgraphs that appear in disassortative networks are more dense than reliable subgraphs from assortative networks. 
We also report the maximum derivative of $R(x)$ with respect to $x$ for GNM graphs in Figure \ref{GNMdrdx}. Clearly, a small change in $x$, i.e.\ $x+dx$, increases the network reliability of graphs with fewer triangles more than that of graphs with more triangles. The influence of assortativity on the maximum derivative of $R(x)$ is more noticeable in assortative graphs than in disassortative graphs. 

\subsubsection{Evaluation of average reliability} % and critical point}

\begin{figure*}
\begin{center}
\subfigure[$\langle R(x) \rangle$ for disassortative GNM graphs $A<0$]{\label{GNMaverxA}\includegraphics[height=5.5cm,width=8.0cm]{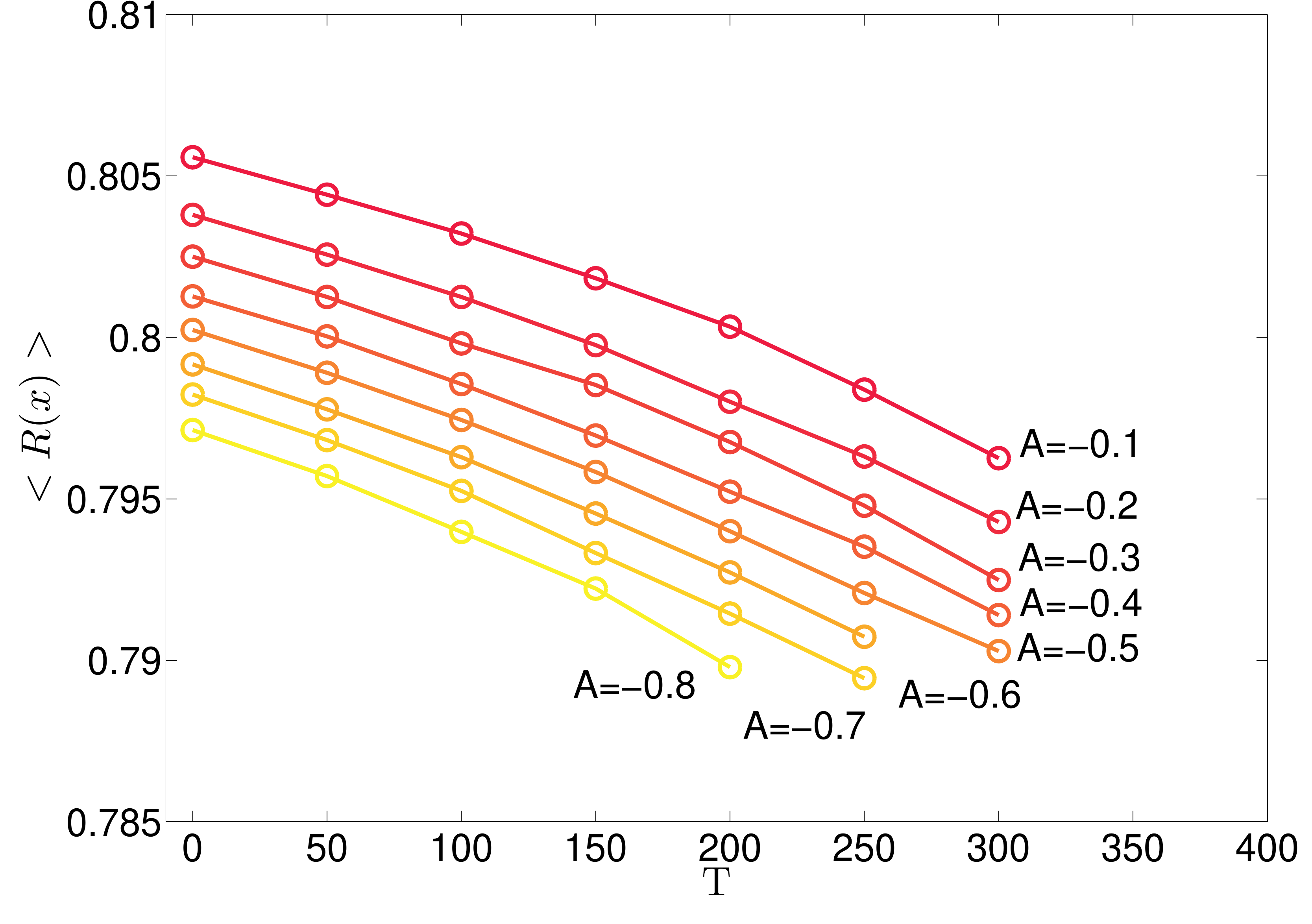}}
\subfigure[$\langle R(x) \rangle$ for neutral and assortative GNM graphs $A\geq 0$]{\label{GNMaverxB}\includegraphics[height=5.5cm,width=8.0cm]{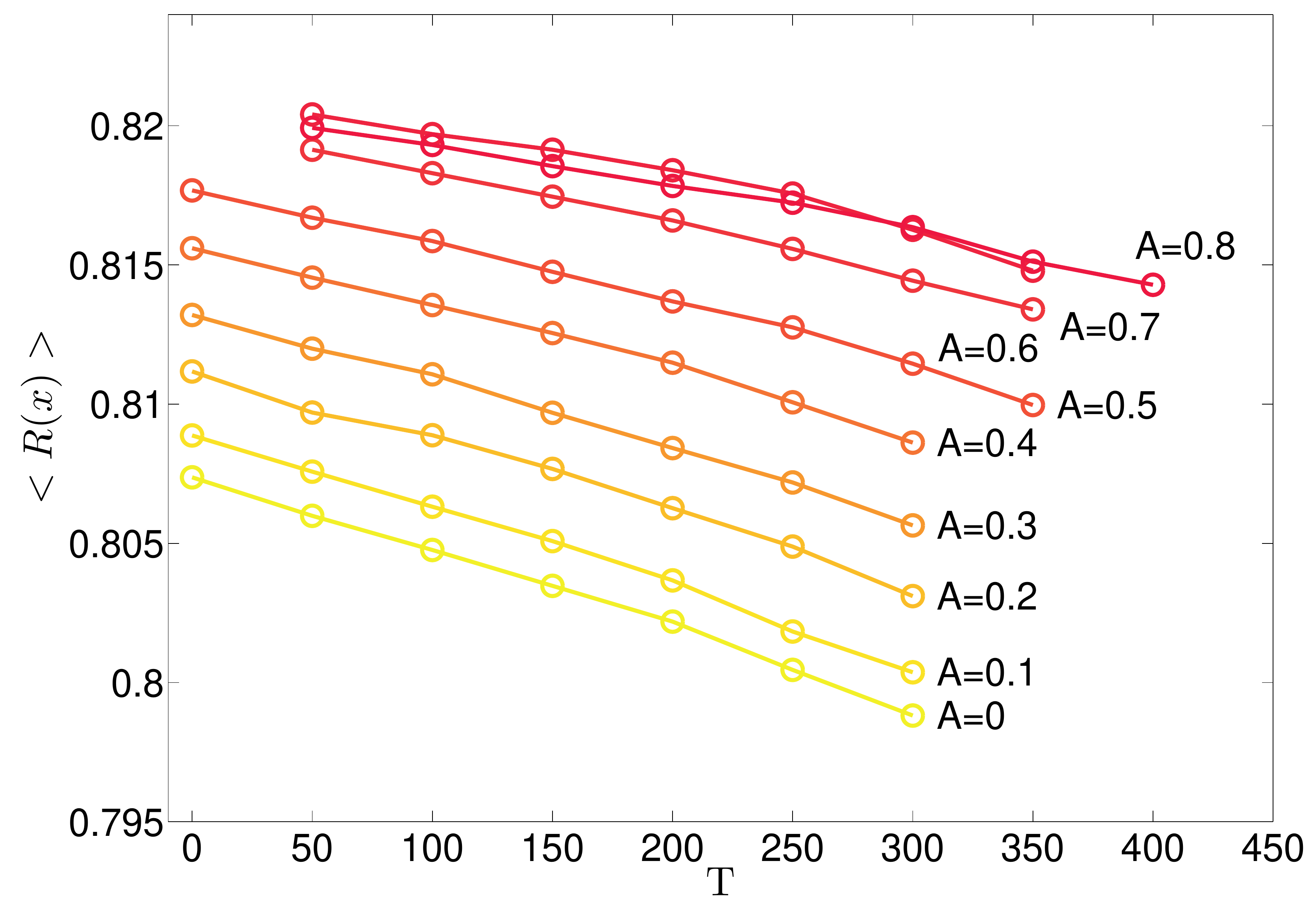}}
\subfigure[Critical point $x_{c}$ for disassortative GNM graphs $A<0$]{\label{GNMcpA}\includegraphics[height=5.5cm,width=8.0cm]{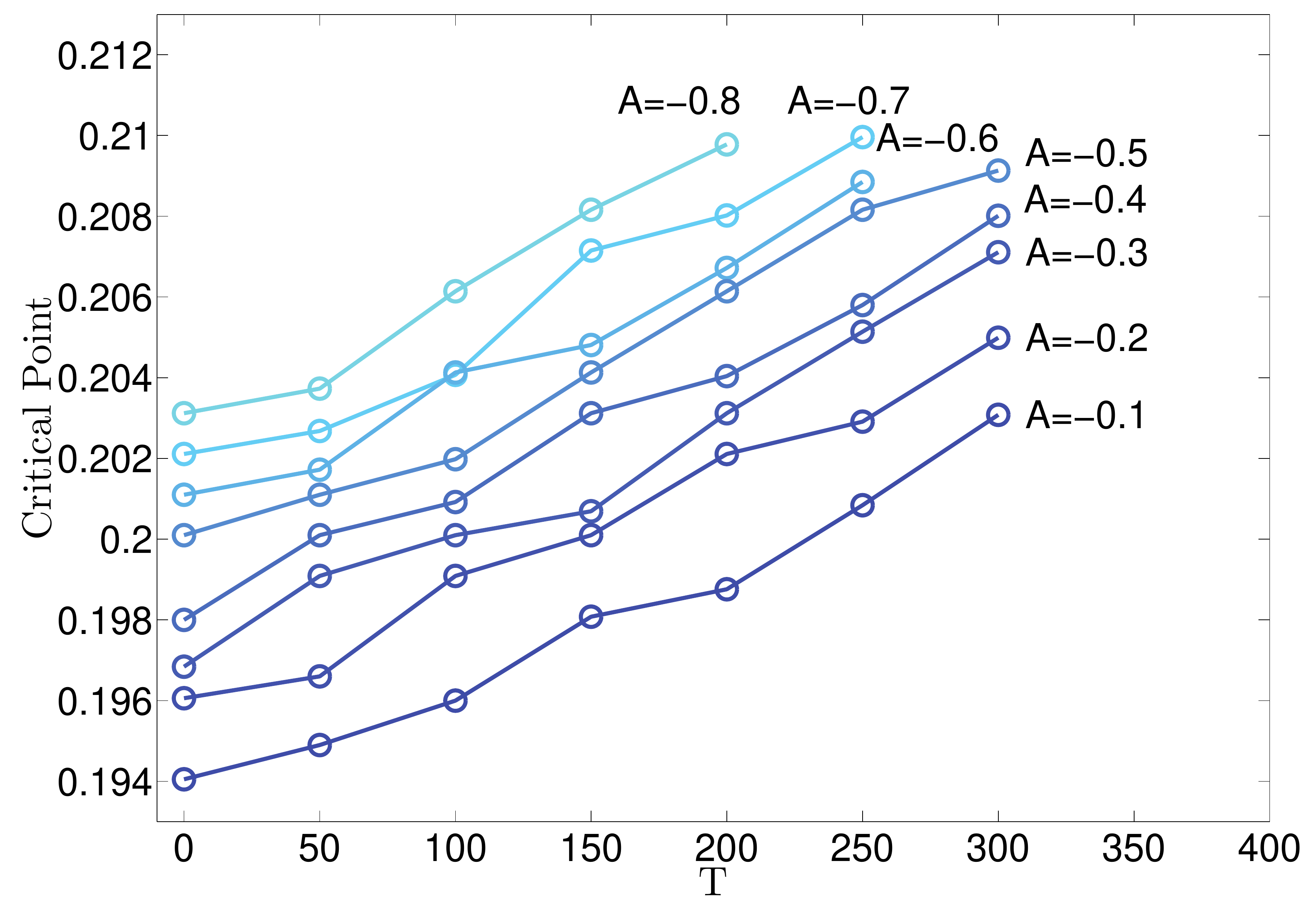}}
\subfigure[Critical point $x_{c}$ for neutral and assortative GNM graphs $A\geq 0$]{\label{GNMcpB}\includegraphics[height=5.5cm,width=8.0cm]{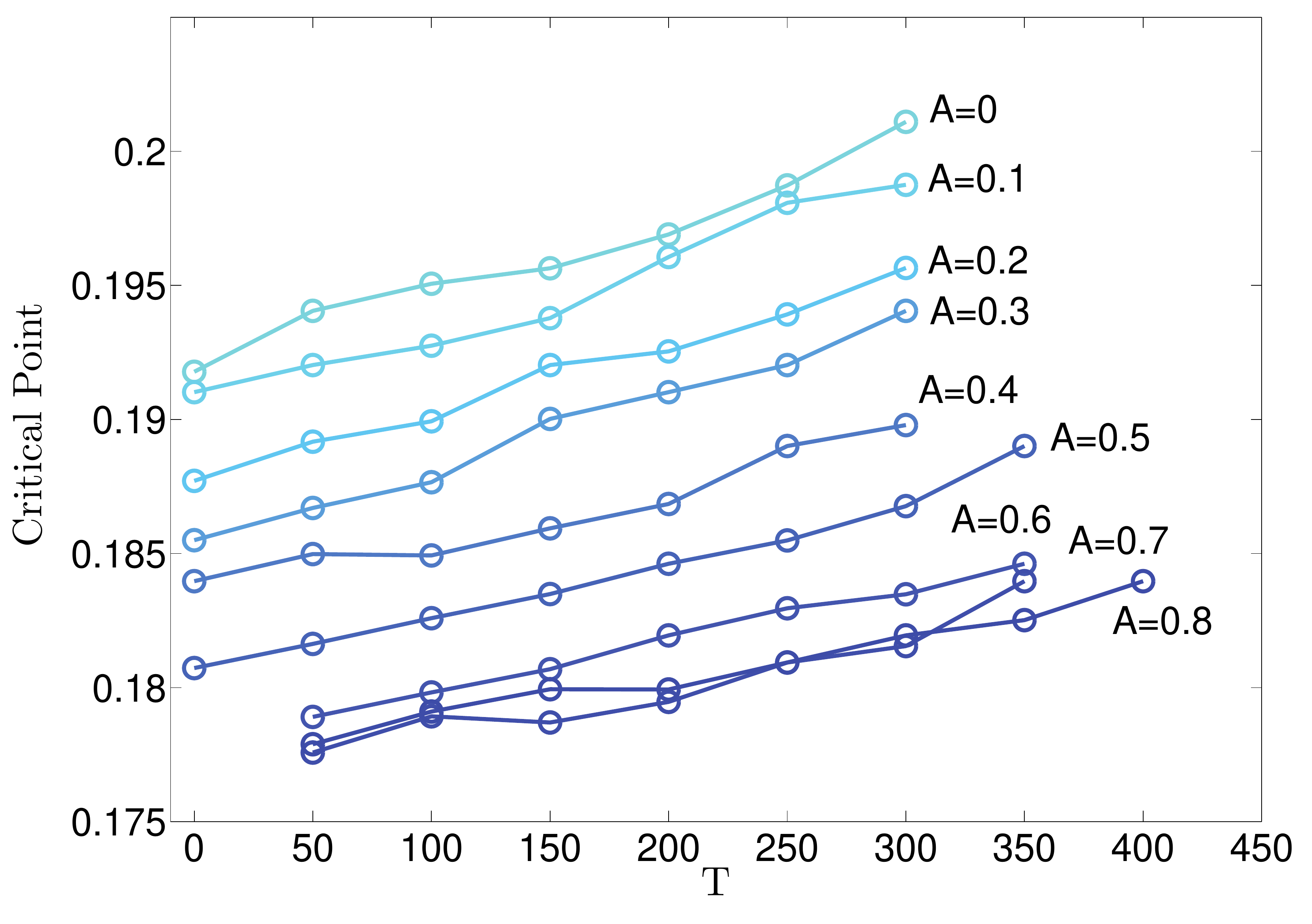}}
\end{center}
\caption{The average reliability $\langle R(x) \rangle$ and the critical points for disassortative, neutral and assortative GNM graphs under an AR-$\alpha$ reliability rule with $\alpha$ = 0.2.}
\label{fig:GNMaverxcp}
\end{figure*}

%We focus our numerical evaluation for network reliability on AR-$\alpha$ rule with $\alpha = 0.2$.
Figures \ref{GNMaverxA} and \ref{GNMaverxB} show the average reliability for GNM networks with negative and neutral or positive assortativity, respectively.  
We first analyze the influence of assortativity and triangles on the reliability independently. 
%{\bf 1) 
\begin{itemize}
\item Effect of triangles on reliability: Network reliability decreases as the number of triangles increases for any assortativity. For the AR-$\alpha$ reliability rule, creating a triangle in an unreliable subgraph by adding a new edge does not make the subgraph reliable since the newly added edge does not increase number of vertices in any connected component. However, if the newly added edge connects a vertex that belongs in one component with a vertex in another component, the probability that the overall subgraph is reliable increases. 

%To understand this phenomenon, assume there is a subgraph that is not reliable and that does not have any triangles. Now suppose that a new edge is added to the subgraph and that it creates one or more triangles. In this scenario, the subgraph remains unreliable since the newly added edge only connects two vertices that already belonged to an unreliable subgraph. However, if the newly added edge connects a vertex that belongs to the unreliable subgraph with another vertex that did not previously belong to the subgraph, the probability that the subgraph becomes reliable increases since the size of the subgraph increases by adding a new vertex and the subgraph may now meet the AR-$\alpha$ reliability rule. 
%{\bf 2) 
\item Effect of assortativity on reliability: The more assortative the network is, the more reliable the network is. We know that reliable subgraphs have lower edge density for assortative graphs than for disassortative graphs i.e. $k_{min}^{assort} < k_{min}^{disassort}$ and $k_{max}^{assort} < k_{max}^{disassort}$. Thus, $\langle P_{k}^{assort} \rangle > \langle P_{k}^{disassort} \rangle$. Consequently, using Eq. \ref{eq:averxpk}, $\langle R(x)^{assort} \rangle$ is larger than $\langle R(x)^{disassort} \rangle$.
\end{itemize} 
In contrast to \cite{Newman:03}, assortative graphs do not always have many cliques. 
Therefore, we analyze the combined effect of the number of triangles and assortativity on the reliability using six distinct combinations of graph properties: \\
%\begin{enumerate}
%\item Assortative graphs with few triangles:
{\bf 1) Assortative graphs with few triangles}:
High degree vertices have high degree neighbors. 
However, these vertices are
not interconnected and hence do not form cliques.
Therefore, reliable subgraphs are weakly locally
connected. It is hard for a reliable subgraph to percolate among only high-degree vertices
because the edge density is lower for the subgraph
containing high degree vertices than for the graph as a whole. Therefore, reliable subgraphs expand
across not only high degree vertices but
also low degree vertices. Due to the assortative property, the majority of vertices will have high degree. Thus, only a small
number of edges is required for a reliable subgraph
to appear. \\
%\item 
{\bf 2) Assortative graphs with many triangles}: 
The majority of edges are used to create triangles among vertices with similar node degrees. In
other words, vertices with similar  degrees form
weakly interconnected cliques. 
Reliable subgraphs appear in cliques with high degree vertices due to their large
edge density. Because the cliques are highly locally
connected, the number of edges in a reliable subgraph
is larger for assortative graphs with large number
of triangles than for assortative graphs with small
number of triangles. In addition, because cliques
are only weakly interconnected, it is hard
for a reliable subgraph to expand outside the clique. \\
%\item 
{\bf 3) Neutral graphs with few triangles}: With equal probability, a randomly selected edge connects vertices with similar degrees or vertices with different degrees. High degree vertices are weakly connected and the subgraph containing them has low edge density. Being neutral and having few triangles in the graph, a reliable subgraph expands across vertices with a wide range of degrees. Therefore, many edges are required to increase the edge density of a subgraph to become reliable. Thus, a reliable subgraph requires more edges for neutral graphs than for assortative graphs, if they both have few triangles.  \\
%\item 
{\bf 4) Neutral graphs with many triangles}: Many triangles exist in the graph without composing cliques. Because the graph is neutral and because triangles do not increase the reliability of graphs, the number of edges needed for a subgraph to be reliable and to expand across the graph is larger for graphs with many triangles than for graphs with few triangles. \\
%\item 
{\bf 5) Disassortative graphs with few triangles}: 
Vertices with different node degrees are connected but do not form cliques. 
Thus, subgraphs with larger edge density than that of the graph as a whole exist with many edges and vertices. 
Consequently, in contrast to reliable subgraphs that appear with fewer edges in assortative graphs with few triangles, reliable subgraphs appear with many edges from high density subgraphs. \\
%Thus, number of edges in a reliable subgraph is large for disassortative graphs than for assortative graphs.
%\item 
{\bf 6) Disassortative graphs with many triangles}: 
Vertices with different node degrees are connected together forming triangles. As discussed
above, in finite graphs, triangles do not increase the reliability of graphs.
%\end{enumerate}

\subsection{Scale-free-like graphs}
\begin{figure*}
\begin{center}
\subfigure[$\langle R(x) \rangle$ for disassortative SFL graphs $A<0$]{\label{SFLaverxA}\includegraphics[height=5.5cm,width=8.0cm]{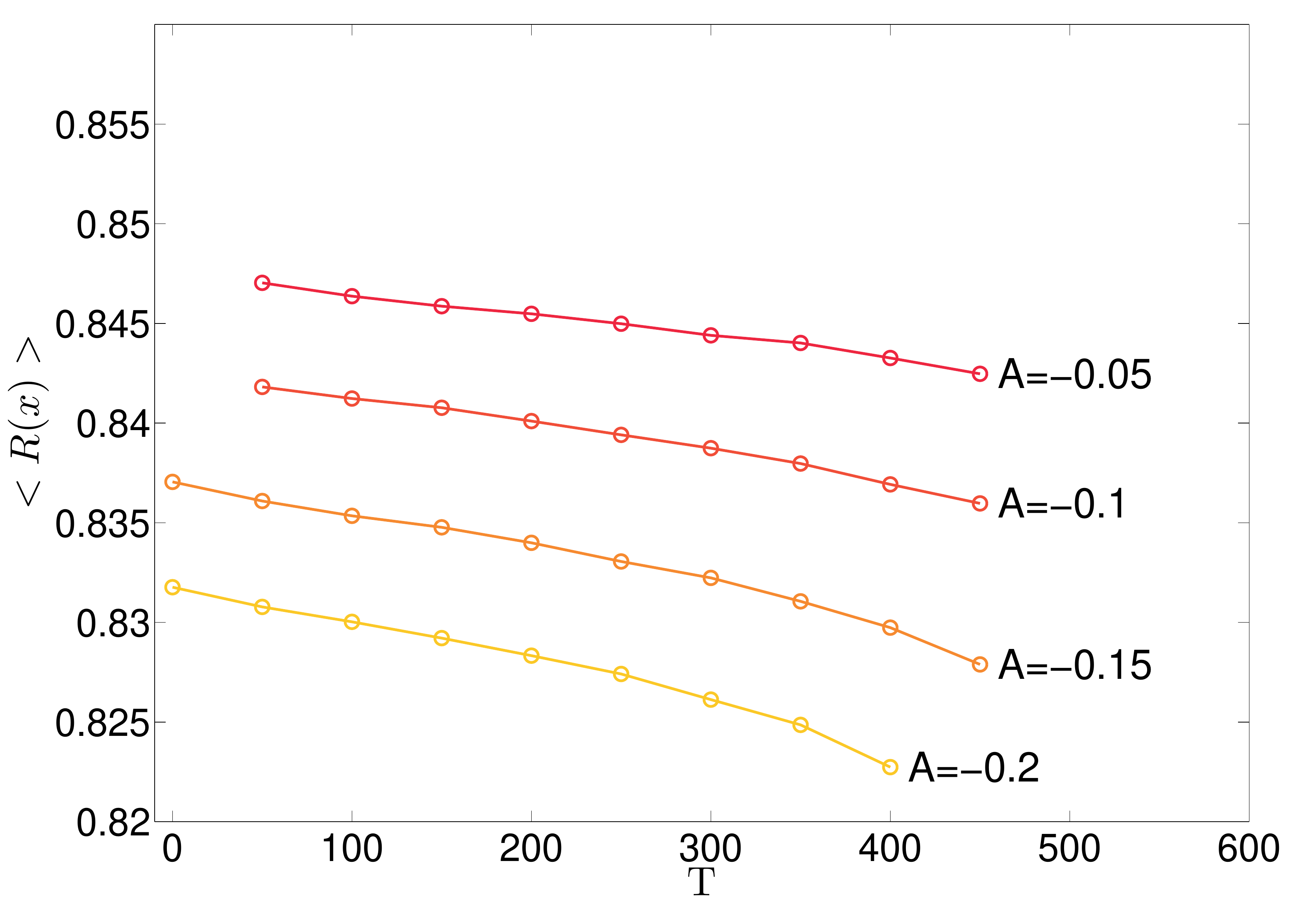}}
\subfigure[$\langle R(x) \rangle$ for neutral and assortative SFL graphs $A\geq 0$]{\label{SFLaverxB}\includegraphics[height=5.5cm,width=8.0cm]{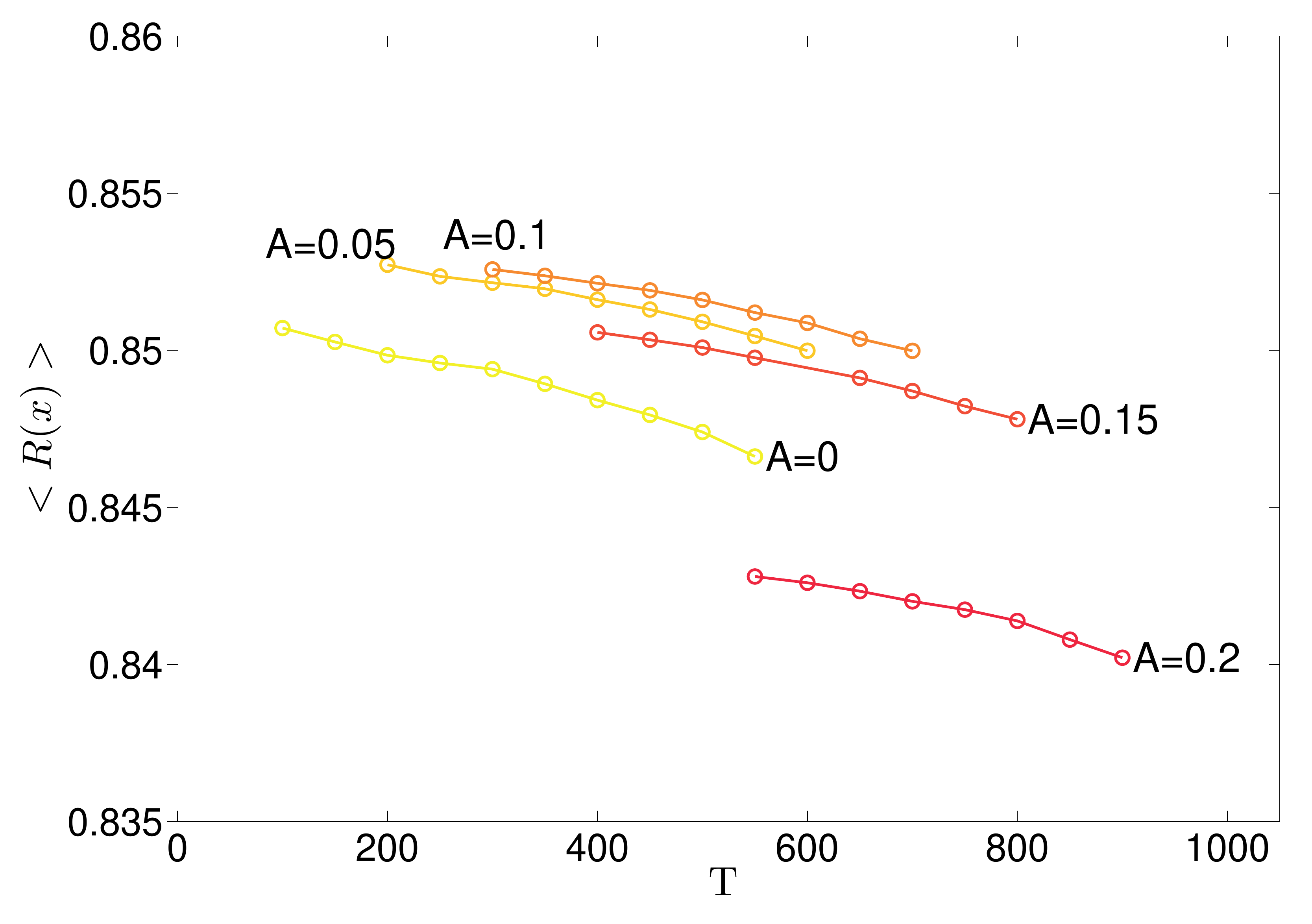}}
\subfigure[Critical point for disassortative SFL graphs $A<0$]{\label{SFLcpA}\includegraphics[height=5.5cm,width=8.0cm]{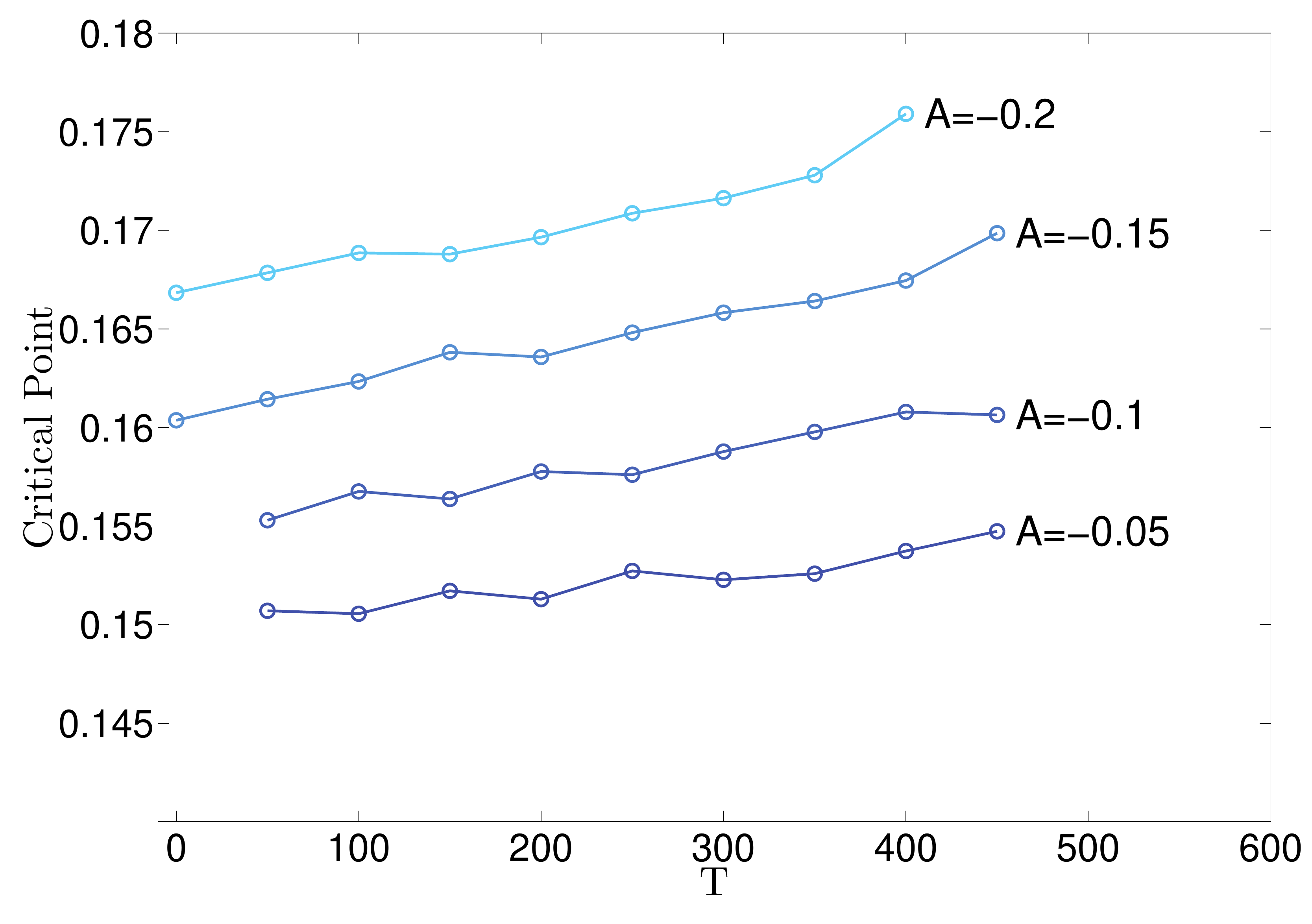}}
\subfigure[Critical point for neutral and assortative SFL graphs $A\geq 0$]{\label{SFLcpB}\includegraphics[height=5.5cm,width=8.0cm]{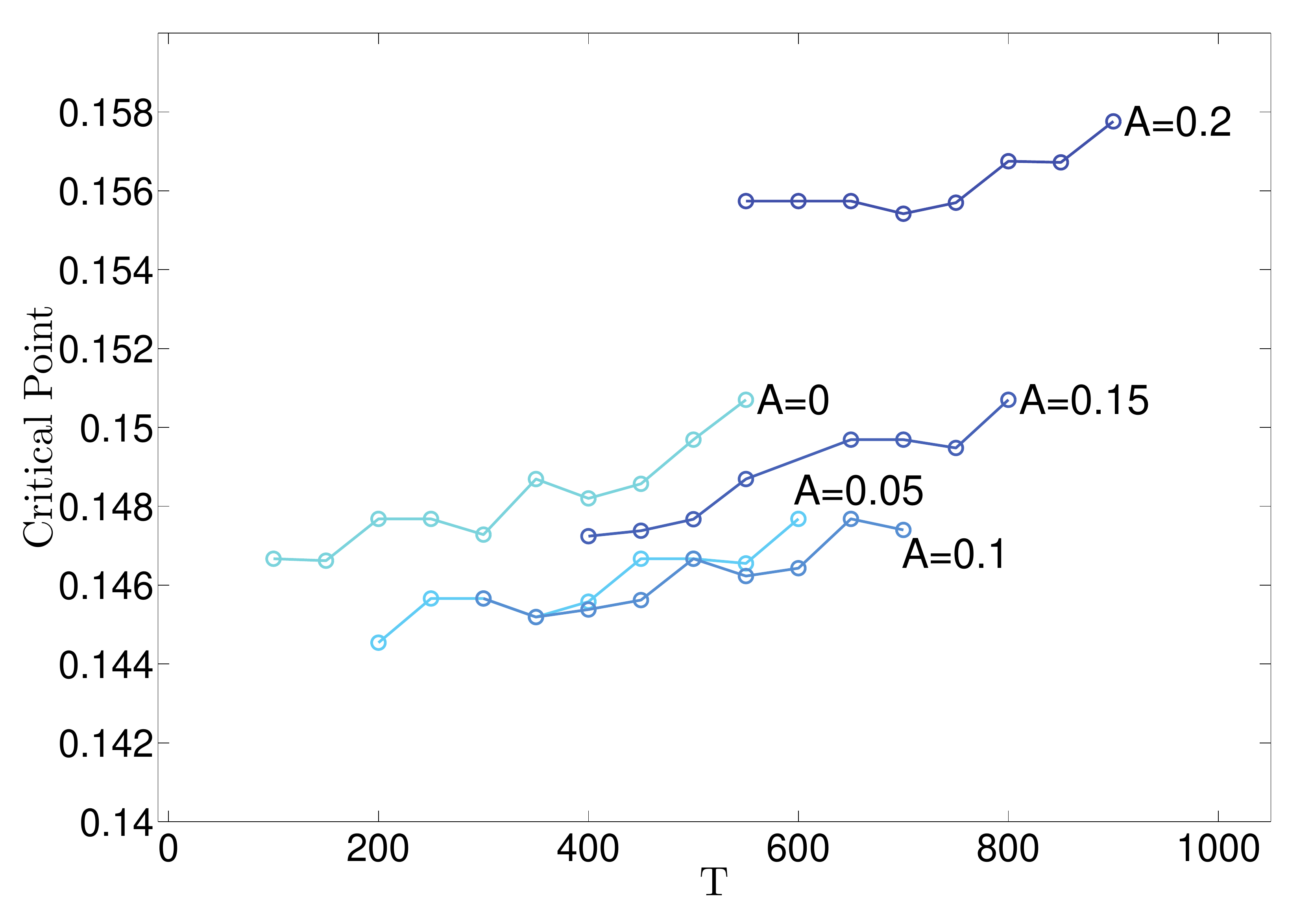}}
\end{center}
\caption{The average reliability $\langle R(x) \rangle$ and the critical points $x_{c}$ for disassortative, neutral and disassortative SFL graphs under an AR-$\alpha$ reliability rule with $\alpha$ = 0.2.}
\label{fig:SFLaverxcp}
\end{figure*}

\begin{figure*}
\begin{center}
\subfigure[GNM graphs]{\label{GNMdrdx}\includegraphics[height=5.5cm,width=8.0cm]{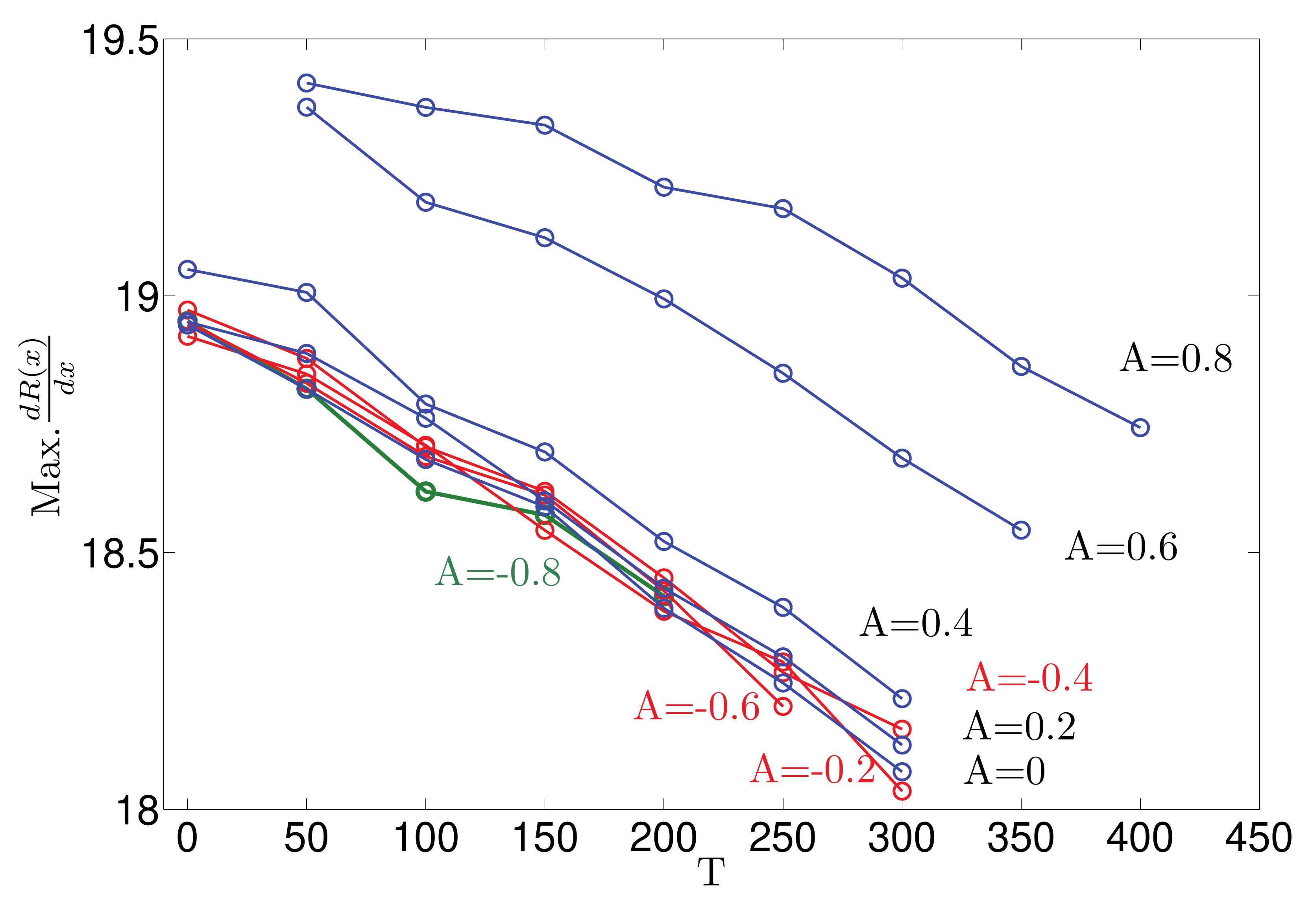}}
\subfigure[SFL graphs]{\label{SFLdrdx}\includegraphics[height=5.5cm,width=8.0cm]{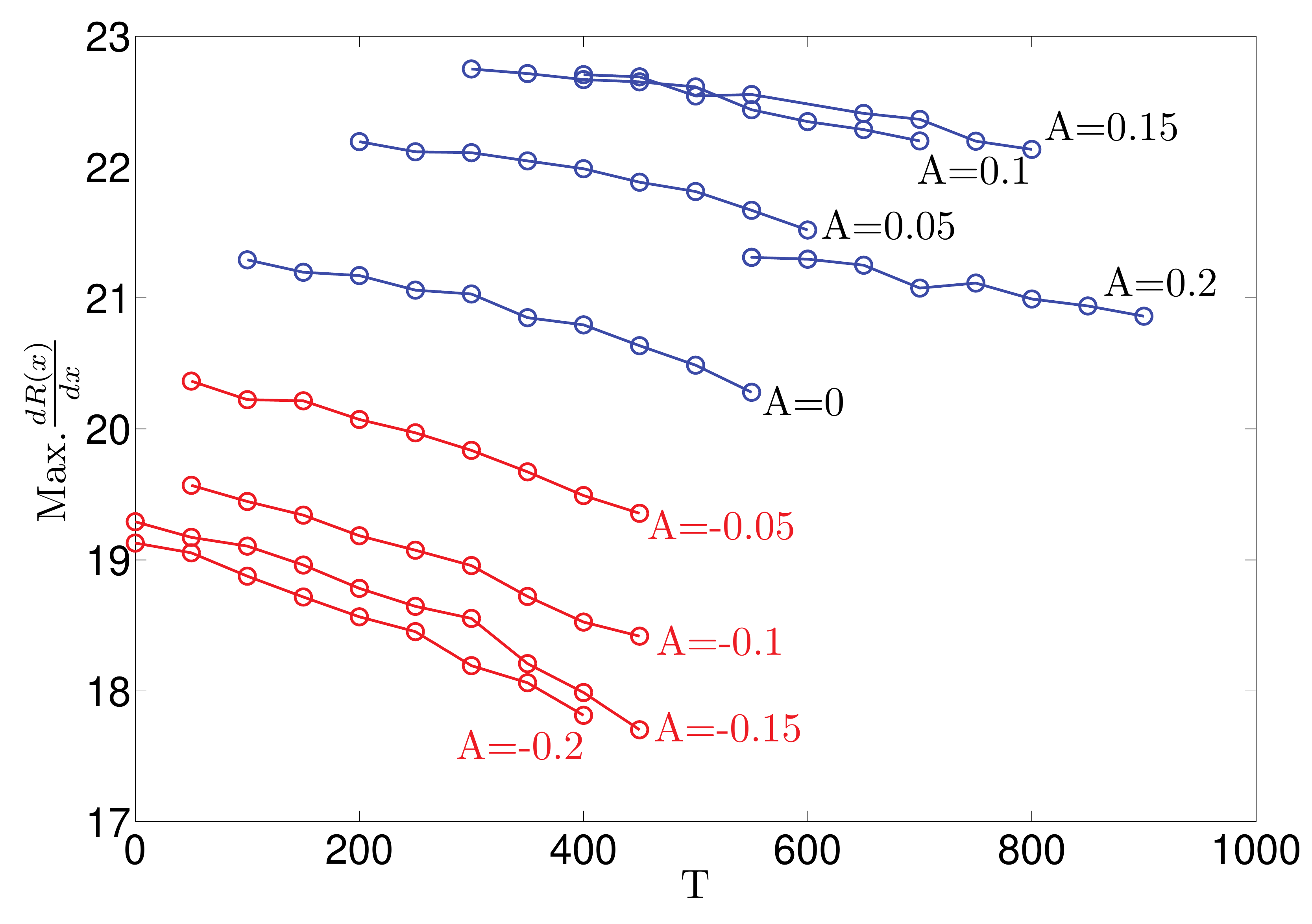}}
%\subfigure[$\frac{dR(x)}{dx}|_{x=critical \ point}$ for disassortative SFL graphs $A<0$]{\label{}\includegraphics[height=5.5cm,width=8.0cm]{.pdf}}
%\subfigure[$\frac{dR(x)}{dx}|_{x=critical \ point}$ for neutral and assortative SFL graphs $A\geq 0$]{\label{}\includegraphics[height=5.5cm,width=8.0cm]{.pdf}}
\end{center}
\caption{ Peak value of the derivative of $R(x)$ for disassortative (red and green), neutral and assortative (blue) GNM and SFL graphs under an AR-$\alpha$ reliability rule with $\alpha$ = 0.2.}
\label{fig:GNMSFLdrdx}
\end{figure*}

%For SFL graphs, both $k_{min}$ and $k_{max}$ increase as the assortativity increases for assortativity range $A>0.1$. 
Results obtained from SFL graphs are in agreement with results from GNM graphs except for assortativity $A>0.1$. For assortativity increases above 0.1, $k_{min}$ and $k_{max}$ increase as shown in Figure \ref{SFLkminkmax}, $\langle R(x) \rangle$ decreases and $x_{c}$ increases as shown in Figure \ref{fig:SFLaverxcp}, and the derivative of reliability at critical point decreases as shown in Figure \ref{SFLdrdx}. Thus, the edge density of reliable subgraphs is larger for SFL graphs with assortativity $A>0.1$ than for neutral SFL graphs. 
To understand this phenomenon, note that SFL graphs with near-maximal assortativity tend to have large number of triangles, because vertices with similar degrees create cliques. These cliques represent communities with vertices that are strongly connected, while different communities are weakly interconnected. Thus, the number of communities decreases~\cite{PSchumm:10} and approaches the number of distinct degree values as assortativity increases for highly assortative SFL graphs. Due to the degree distribution of SFL graphs, the majority of lowest degree vertices belong to a single community. The edge density is lower for this community than for the graph as a whole. Conversely, the communities of high degree vertices contain only a few vertices. 
Therefore, for reliable subgraphs to appear in communities with high edge density, the reliable subgraphs have to extend across different communities that are weakly interconnected. Consequently, a large number of edges is required to obtain reliable subgraphs from highly assortative SFL graphs. This result causes the critical point to increase with assortativity leading to a decrease in the average reliability. 

\subsection{Network reliability and scaling}

We study the effect of graph size by evaluating the reliability on GNM graphs with fixed average node degrees and sizes $V$, $2V$ and $4V$. Three different assortativity values are used, while the number of triangles is held constant at 100. The results are summarized in Table \ref{scaling}. Let $k'$ be the normalized number of edges with respect to the total number of edges in the graph, e.g. $k_{min}'=\frac{k_{min}}{4E}$ for graphs with $4V$ vertices and $4E$ edges. We observe that the average reliability, $k_{min}'$ and maximum derivative increase as the graph size increases, while $k_{max}'$, $x_{c}$ and $k_{max}'-k_{min}'$ decrease as the graph size increases. In addition, results show that the derivative of the reliability with respect to $x$ diverges for larger graph sizes. In other words, the transition from $R(x)=0$ to $R(x)=1$ becomes sharper for large graphs than for small graphs. Consequently, at the thermodynamic limit, $k_{max}'- k_{min}'$ converges to 0 i.e. $k_{min}'$ and $k_{max}'$ reach their convergence value $k_{therm}'$. Thus, $\langle R(x) \rangle$ and $x_{c}$ converge to $1-k_{therm}'$ and $k_{therm}'$, respectively. Therefore, network reliability moves toward a sharp transition for infinite size systems, reflecting a first order phase transition from a region of unreliable subgraphs on one side to a region with only reliable subgraphs on the other side.

% In addition, the phase transition takes place for smaller value of critical point. Therefore, $<R(x)>$ increases as the graph size increases. 
%Moreover, the transition from $P_{k}=0$ to $P_{k}=1$ becomes sharper for large graphs.

\begin{table}%[H] add [H] placement to break table across pages
\caption{\label{scaling} Evaluation of average reliability $\langle R(x) \rangle$, $k_{min}'$, $k_{max}'$, derivative of reliability at critical point and the critical point $x_{c}$ for GNM graphs with different graph sizes $V$, $2V$ and $4V$. Each graph has assortativity $A$ =-0.85, 0 and 0.85 and number of triangles $T=100$.}
\begin{ruledtabular}
\begin{tabular}{|l|c|c|c|}
%Lines of table here ending with \\
$A=-0.85$ & $V$ & $2V$ & $4V$ \\ \hline
$\langle R(x) \rangle$ & 0.7935 & 0.7950 & 0.7972 \\ \hline
$k_{min}'$ & 0.1522 & 0.1623 & 0.1767 \\ \hline
$k_{max}'$ & 0.2550 & 0.2440 & 0.2349 \\ \hline
$k_{max}'-k_{min}'$ & 0.1028 & 0.0817 & 0.0582 \\ \hline
max $\frac{dR(x)}{dx}$ & 18.6515 & 25.4679 & 38.7695 \\ \hline
$x_{c}$ & 0.2066 & 0.2046  & 0.2021 \\ \hline\hline

$A=0$ & $V$ & $2V$ & $4V$ \\ \hline
$\langle R(x) \rangle$ & 0.8049 & 0.8066 & 0.8067 \\ \hline
$k_{min}'$ & 0.1391 &  0.1563 & 0.1641 \\ \hline
$k_{max}'$ &  0.2460 & 0.2319 & 0.2228 \\ \hline
$k_{max}'-k_{min}'$ & 0.1069 & 0.0756 & 0.0587 \\ \hline
max $\frac{dR(x)}{dx}$ & 18.7144 & 25.5718 & 35.2221 \\ \hline
$x_{c}$ & 0.1945 & 0.1925 & 0.1928 \\ \hline\hline

$A=0.85$ & $V$ & $2V$ & $4V$ \\ \hline
$\langle R(x) \rangle$ & 0.8173 & 0.8228 & 0.8311 \\ \hline
$k_{min}'$ &  0.1270 &  0.1462 &  0.1447 \\ \hline
$k_{max}'$ &  0.2429 & 0.2172  & 0.2016 \\ \hline
$k_{max}'-k_{min}'$ & 0.1159 & 0.0710 & 0.0569 \\ \hline
max $\frac{dR(x)}{dx}$ & 18.9097 & 27.5449 & 38.7695 \\ \hline
$x_{c}$ & 0.1804  & 0.1759   & 0.1680 \\
\end{tabular}
\end{ruledtabular}
\end{table}

\section{Conclusion and future work}

The classical concept of network reliability provides a rich theoretical basis, supported by computational estimation procedures, to study the effect of structural properties on diffusion dynamics. 
% We have derived two equivalent representations for network reliability and 
We have highlighted various features of reliability that provide useful characterizations of graph structure, e.g.\ the minimum and maximum number of edges needed to obtain reliable subgraphs, the average reliability and the critical point. 
We have created and made widely available a library of graphs with carefully controlled structural properties, i.e.\ assortativity-by-degree and triangles. 

Simulation results for Erd\H{o}s-R\'{e}nyi and scale-free-like random graphs in this library reveal that increasing the  assortativity and number of triangles has opposite effects on the probability that an epidemic outbreak will achieve an average attack rate of 20\%. We found that the required number of edges decreases as the degree assortativity increases; however, the required number of edges increases as the number of triangles increases. 
In addition, average network reliability increases as the degree assortativity increases but decreases as the number of triangles increases. 
Moreover, the critical point decreases and the derivative of reliability at critical point diverges as the degree assortativity increases, while the opposite is true for increasing number of triangles. In contrast to assortative GNM graphs, network reliability decreases as assortativity increases for assortative SFL graphs.   Furthermore, we have demonstrated that the transition from unreliable subgraphs to reliable subgraphs behaves as expected.
% the effect of graph size on network reliability showing that the transition  becomes sharper for large graphs than for small graphs. Thus, mean reliability increases as the graph size increases. \newline

Obviously, there are many avenues for future work in this area, such as studying the relationship between reliability and other common graph statistics. 
In a companion paper, we show the relationship between network reliability and statistical physics and we demonstrate the power of reliability for reasoning about graph structure using the overlaps of structural
motifs. 
We also introduce a new measure of centrality -- similar to betweenness but more closely tailored to specific dynamics -- and use it to compare graphs.
To extend the application of network reliability to epidemiology, we will use
reliability to characterize large, realistic social networks and the effect of changes brought about by outbreak control interventions.

\begin{acknowledgments}
We thank our external collaborators and members of the Network Dynamics and Simulation Science Laboratory (NDSSL) for their suggestions and comments, particularly M. Marathe and A. Vullikanti.  This  work has been partially supported by DTRA R\&D Grant HDTRA1-0901-0017, and DTRA CNIMS Grant HDTRA1-07-C-0113. Research reported in this publication was supported by the National Institute of General Medical Sciences of the National Institutes of Health under NIH MIDAS Grant 2U01GM070694-09. The content is solely the responsibility of the authors and does not necessarily represent the official views of the National Institutes of Health or DTRA.
\end{acknowledgments}

\appendix
%\section{\label{sec:Nkconstraints}Constraints on coefficients $R_k$ and $P_k$}
%
%%\subsection{Constraints on $R_k$ and $P_k$}
%$R_k$ is an integer in the range $[0, {E \choose k}]$.
%For a coherent reliability rule, $P_k$ is monotonic non-decreasing.
%Combining these two constraints, we have
%\begin{eqnarray}
%P_{k+1} &=& P_k (1+\Delta_k), \quad {\rm where } \quad \frac{1-P_k}{P_k} \ge \Delta_k \ge 0\nonumber  \\
%R_{k+1} {E \choose k} &=& R_k {E \choose k+1} (1+\Delta_k) \nonumber \\
%R_{k+1} &=& R_k \frac{E-k}{k+1} (1+\Delta_k) \nonumber \\
%R_{k+1} &=& R_k \frac{(E-k)(1+\Delta_k)}{k+1},
%\end{eqnarray}
%whence $k+1$ must divide $R_k (E-k)(1+\Delta_k)$, or
%$\Delta_k = \alpha_k (k+1) - (R_k (E-k)) {\rm mod} (k+1)$, where $\alpha_k$ is an integer $\ge 1$.
%Obviously, if $P_k = 0$, then $P_{k+1}$ must be an integer multiple of ${E \choose k+1}^{-1}$.

\section{Implementing \AsUp, \AsDown, \TriUp, and \TriDown}
\begin{figure}
\includegraphics[width=9cm]{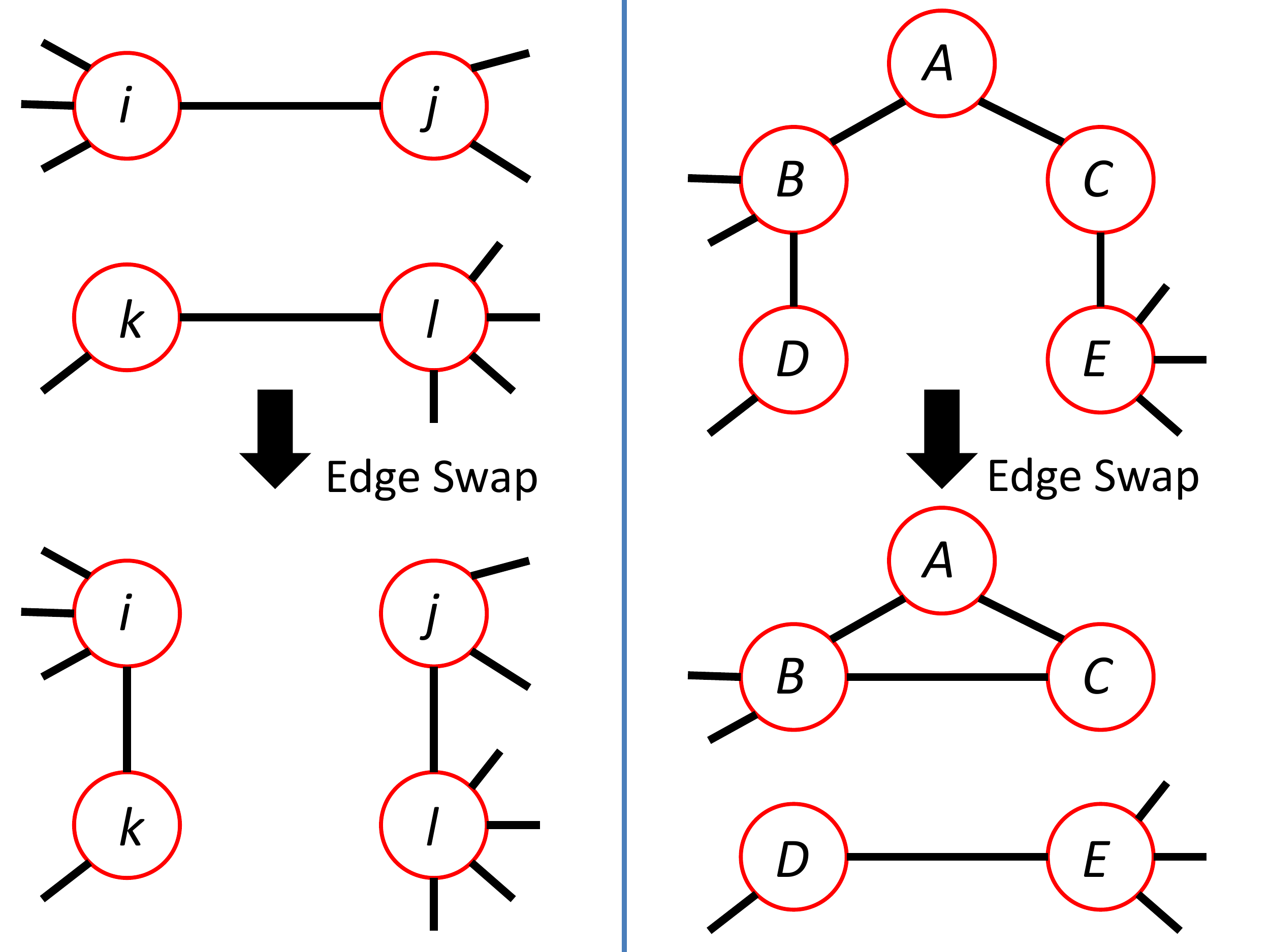}
\caption{A degree distribution-preserving edge swap. The four vertices shown are connected by the edge stubs shown here to the rest of the graph, which is not shown. (left panel) Assortativity-changing edge swap:
In this example, $d_i=4$, $d_j=3$, $d_k=2$, and $d_l=5$.
Hence $d_id_j + d_kd_l = 22$ and $d_id_k + d_jd_l = 23$ so the graph at the bottom has a higher assortativity than the one at the top \cite{Mieghem:10,Mieghem:12}.
If the graph at the top is connected, the graph at the bottom will also be connected if there is a path from $i$ to $j$.
In that case, this will be an acceptable edge swap for $A^+$, but not for $A^-$. (right panel) Assortativity-preserving and triangle-changing edge swap: 
$B$ and $D$ have no neighbors in common, nor do $C$ and $E$.
In this example, $d_B=d_E=4$ and $d_C=d_D=2$.
Hence $d_Bd_D+ d_Cd_E = 40=d_Bd_C+ d_Dd_E$ so the graph at the top has the same assortativity as the one at the bottom.
If the graph at the top is connected, the graph at the bottom will also be connected if there is a path from $B$ to $D$.
In that case, the swap from top configuration to bottom configuration will be an acceptable one for $T^+$;
 the swap in the opposite direction will be acceptable for $T^-$, since the swap cannot disconnect the graph.
This swap changes the number of triangles in the graph by at least one -- more, if $B$ and $C$ have any  common neighbors besides $A$. }
\label{fig:edgeswap}
\end{figure}

We implement the $A^{\pm}$ and $T^{\pm}$ operators using edge swapping, or graph rewiring. %\cite{Strogatz, NDSSL, ERGMs}.
We choose candidate edges uniformly at random from among all the edges in the graph and, if the candidates meet certain constraints, swap them.
Figure~\ref{fig:edgeswap} illustrates the swaps involved.

Specifically, given a graph $G$ defined by edge set ${\cal E}$, which includes edges $(i,j)$ and $(k,l)$, the operator $A^+_{ij,kl}$ (resp. $A^-_{ij,kl}$)
returns either the same graph $G$ or a new graph $G'$ with the edge set ${\cal E} - \{(i,j), (k,l)\} + \{(i,k), (j,l)\}$, whichever increases (resp. decreases) the assortativity.
That is, $A^+_{ij,kl}(G) = \argmax_{G, G'} a(g)$.
We check the constraints that $i,j,k,l$ are all distinct, that the edges $(i,k)$ and $(j,l)$ do not already exist, and that the graph $G'$ remains connected.
(This last constraint can be checked by ensuring that the pairs of vertices originally connected by edges are in the same component of $G'$.).
Since this edge swap does not change the degree of the affected vertices, the direction of the change in assortativity is easily computed by comparing the values $d_id_j + d_kd_l$ and $d_id_k + d_jd_l$. 

The triangle operators must satisfy more constraints, both because they are intended to maintain the assortativity invariant and because triangles are less local than edges.
In this case, we randomly choose a vertex $A$, and randomly pick two of its neighbors, $B$ and $C$, that are not connected by an edge.
As illustrated in the right panel of Figure~\ref{fig:edgeswap}, we find a neighbor $D$ of $B$ that is not $A$ or $C$, has the same degree as $C$, and has no neighbors in common with $B$.
I.e.\ the edge $(B,D)$ is not a part of any triangle.
We repeat this, replacing vertex $B$ with $C$ to find $E$, with the additional constraint that $E$ is not a neighbor of $D$.
Then we swap edges $(B,D)$ and $(C,E)$ for edges $(B,C)$ and $(D,E)$.
By construction, this does not change the assortativity, but it creates at least one more triangle than was present before, namely $(A,B,C)$.

The $T^-$ operator accomplishes the swap from the bottom of the right panel of Figure~\ref{fig:edgeswap} to the top of the right panel. 
We first find vertices $(A,B,C)$ that form a triangle.
Then we find an edge $(D,E)$, such that 1) $D$ and $E$ are both different from $A$, 2) $D$ and $E$ are not neighbors of $A$, $B$, or $C$, 3) $E$ has the same degree as $B$, 4) $D$ has the same degree as $C$, 5) $B$ and $D$ have no common neighbor and 6) $C$ and $E$ have no common neighbor. Then, as usual, we swap edges and test for connectivity in the new graph.

%\subsection{Manipulating assortativity and clustering}

\bibliography{PREpaper}

\end{document}